\documentclass[12pt]{article}

\usepackage{a4,graphicx,color}
\renewcommand{\title}[1]{
\begin{center} \Large \bf #1 \end{center}
}

\renewcommand{\author}[2]{
 \begin{center} #1  \vspace{3mm} \\
  #2 \\
 \end{center}
\addvspace{\baselineskip}
}

\makeatletter
\renewenvironment{thebibliography}[1]
         {\section*{References}\addcontentsline{toc}{section}{References}%
          \frenchspacing
          \begin{list}{[\arabic{enumi}]}
         {\usecounter{enumi}\parsep=2pt\@plus\p@\topsep 0pt
         \settowidth{\labelwidth}{[#1]}
         \leftmargin=\labelwidth\advance\leftmargin\labelsep
         \rightmargin=0pt\itemsep=0pt\sloppy}}{\end{list}}
\makeatother

\usepackage{amssymb}
\usepackage{amsmath}
\usepackage{hyperref}

\def\under#1{\kern.4pt\underline{\kern-.4pt{}#1\kern-.4pt}\kern.4pt}
\definecolor{math1}{rgb}{0.368417, 0.506779, 0.709798}
\definecolor{math2}{rgb}{0.880722, 0.611041, 0.142051}
\definecolor{math3}{rgb}{0.560181, 0.691569, 0.194885}
\definecolor{math4}{rgb}{0.922526, 0.385626, 0.209179}
\definecolor{math5}{rgb}{0.528488, 0.470624, 0.701351}
\definecolor{math6}{rgb}{0.772079, 0.431554, 0.102387}

\usepackage{amsthm}
\newtheorem{thm}{Theorem}[section]
\newtheorem{prop}[thm]{Proposition}

\theoremstyle{definition}
\newtheorem{defn}[thm]{Definition}

\theoremstyle{remark}

\makeatletter
\@addtoreset{equation}{section}

\makeatother

\allowdisplaybreaks[4]

\begin{document}

\baselineskip 5mm

\title{The $\Phi^3_4$ and $\Phi^3_6$ matricial QFT models
have \\ reflection positive 
two-point function}

\author{Harald Grosse${}^1$, Akifumi Sako${}^{1,2}$
and Raimar Wulkenhaar${}^3$}{
${}^1$
Fakult\"at f\"ur Physik, Universit\"at Wien\\
Boltzmanngasse 5, A-1090 Wien, Austria\\[\smallskipamount]
${}^2$  Department of Mathematics,
Faculty of Science Division II,\\
Tokyo University of Science,
1-3 Kagurazaka, Shinjuku-ku, Tokyo 162-8601, Japan\\[\smallskipamount]
${}^3$ {Mathematisches Institut der Westf\"alischen
  Wilhelms-Universit\"at\\
Einsteinstra\ss{}e 62, D-48149 M\"unster, Germany}}

\vspace{1cm}

\footnotetext[1]{harald.grosse@univie.ac.at,
$^2$sako@rs.tus.ac.jp, $^3$raimar@math.uni-muenster.de}

\hspace*{\fill}\emph{dedicated to the memory of Wolfhart Zimmermann (1928--2016)}

\bigskip

\begin{abstract} 
  \noindent We extend our previous work (on $D=2$) to give an exact
  solution of the $\Phi^3_D$ large-$\mathcal{N}$ matrix model (or
  renormalised Kontsevich model) in $D=4$ and $D=6$
  dimensions. Induction proofs and the difficult combinatorics are
  unchanged compared with $D=2$, but the renormalisation -- performed
  according to Zimmermann -- is much more involved. As main result we
  prove that the Schwinger 2-point function resulting from the
  $\Phi^3_D$-QFT model on Moyal space satisfies, for real coupling
  constant, reflection positivity in $D=4$ and $D=6$ dimensions. The
  K\"all\'en-Lehmann mass spectrum of the associated Wightman 2-point
  function describes a scattering part $|p|^2\geq 2\mu^2$ and an
  isolated fuzzy mass shell around $|p|^2=\mu^2$.
\end{abstract}

\section{Introduction}

The Kontsevich model \cite{Kontsevich:1992ti, Witten:1991mn} is of
paramount importance because it elegantly proves Witten's conjecture
\cite{Witten:1990hr} about the equivalence of two approaches to quantum
gravity in two dimensions: the Hermitean one-matrix model
\cite{Brezin:1990rb, Douglas:1989ve, Gross:1989vs} versus the
intersection theory on the moduli space of Riemann surfaces
\cite{Labastida:1988zb, Montano:1988nw, Myers:1989dn}.  The Kontsevich
model is defined by the partition function (we use different notation)
\begin{align}
\mathcal{Z}[E]:=
\frac{\displaystyle 
\int d\Phi \;\exp\Big(-\mathrm{Tr}\big( E \Phi^2
+ \tfrac{\mathrm{i}}{6} \Phi^3\big)\Big)}{
\displaystyle \int d\Phi \;\exp\Big(-\mathrm{Tr}\big( E \Phi^2
\big)\Big)},
\label{Kontsevich}
\end{align}
where the integral is over the space of self-adjoint 
$(\mathcal{N}\times \mathcal{N})$-matrices with Lebesgue measure
$d\Phi$. Perturbative expansion in the `coupling constant'
$\tfrac{\mathrm{i}}{6}$ yields an asymptotic expansion of $\log
\mathcal{Z}[E]$ in terms of rational functions of the eigenvalues
$\{e_i\}$ of the positive self-adjoint matrix $E$.  Kontsevich's main
theorem states that $\log \mathcal{Z}[E]$ is, when viewed as rational function
of $\{e_i\}$, the generating function for the 
intersection numbers.

In a series of papers
\cite{Grosse:2005ig, Grosse:2006qv, Grosse:2006tc}, one of us (H.G.) 
with H.~Steinacker turned the 
Kontsevich model into a quantum field theory on noncommutative Moyal 
space (inspired by the work of Langmann-Szabo-Zarembo 
\cite{Langmann:2003if}). 
The action (\ref{Kontsevich}) was generalised to 
admit a purely imaginary coupling constant 
$\tfrac{\mathrm{i}}{6} \mapsto  \frac{\mathrm{i}}{6} \lambda$, 
and the interest was turned to correlation functions arising from 
the formal measure (\ref{Kontsevich}),
\begin{align}
\big\langle \Phi_{k_1l_1} \cdots
\Phi_{k_Nl_N} \big\rangle := 
\frac{\displaystyle 
\int d\Phi \;\Phi_{k_1l_1} \cdots
\Phi_{k_Nl_N}\,
\exp\Big(-\mathrm{Tr}\big( E \Phi^2 +A \Phi
+ \tfrac{\mathrm{i}}{6} \lambda \Phi^3\big)\Big)}{
\displaystyle \int d\Phi \;\exp\Big(-\mathrm{Tr}\big( E \Phi^2 +A \Phi
+ \tfrac{\mathrm{i}}{6} \lambda \Phi^3\big)\Big)}.
\label{Harald+Harold}
\end{align}
With the usual trick of generating $\Phi_{kl}=
\frac{\partial}{\partial J_{lk}} e^{\mathrm{tr}(\Phi J)}\big|_{J=0}$
under the integral and after absorbing $J,\lambda$ in a redefinition
of $E$ and $\Phi$, the known solution (in the large-$\mathcal{N}$
limit) of the Kontsevich action (\ref{Kontsevich}) allows to formally
derive all correlation functions (\ref{Harald+Harold}). The term
`formally' refers to the fact that the large-$\mathcal{N}$ limit
produces the usual divergences of quantum field theory. Their removal
by appropriate choice, depending on dimension $\{2,4,6\}$, of
parameters in $E,A,\lambda$ was achieved in \cite{Grosse:2005ig,
  Grosse:2006qv, Grosse:2006tc}. Explicit formulae for the
renormalised functions (\ref{Harald+Harold}) at $N\in \{1,2,3\}$ and
genus $g=0$ were given.

In a recent paper \cite{Grosse:2016pob} we transferred the solution
strategy developed by two of us (H.G.+R.W.) for the
$\Phi^4_4$-matricial quantum field theory \cite{Grosse:2012uv} to the
$\Phi^3_2$-Kontsevich model. This approach uses the Ward-Takahashi
identity for the $U(\mathcal{N})$ adjoint action (first employed in
\cite{Disertori:2006nq}) to derive a system of Schwinger-Dyson
equations in which the $N$-point function depend only on $N'$-point
functions with $N'\leq N$. The initial step for the \mbox{Kontsevich} model
is, in a special limit of large matrices coupled with an infinitely
strong deformation parameter, an integral equation solved by Makeenko
and Semenoff \cite{Makeenko:1991ec}. From this point of origin we
explicitly solved all correlation functions.

The present paper extends \cite{Grosse:2016pob} to $D=4$ and $D=6$
dimensions. Similar as in \cite{Grosse:2006qv, Grosse:2006tc} we can
recycle almost everything from two dimensions; only the
renormalisation is more involved. We follow the renormalisation
philosophy advocated by Wolfhart Zimmermann according to which the
theory is defined by \emph{normalisation conditions}, at the physical
energy scale, of a few relevant and marginal couplings. In its
culmination due to Zimmermann \cite{Zimmermann:1969jj}, this BPHZ
renormalisation scheme \cite{Bogoliubov:1957??, Hepp:1966eg, 
Zimmermann:1969jj} completely avoids the use of ill-defined
(divergent) quantities. That the BPHZ scheme extends smoothly to 
non-perturbative renormalisation is our first noticeable message. 

For reasons explained below we are particularly interested in \emph{real}
$\Phi^3$ coupling constant, i.e.\ $\mathrm{i}\lambda\mapsto \lambda$
in (\ref{Harald+Harold}). This is a drastic change! The partition
function does not make any sense for real coupling constants. Our
point of view is to \emph{define} quantum field theory by quantum
equations of motion, i.e. Schwinger-Dyson equations. These equations
can formally be derived from the partition function, but then we
forget the partition function, declare the equations as exact and
construct exact solutions. Whereas the $\Phi^3_6$-Kontsevich model with
imaginary coupling constant is asymptotically free 
\cite{Grosse:2006tc}, our real $\Phi^3_6$-model has 
positive $\beta$-function. But this is not a problem; there is no 
Landau ghost, and the theory remains well-defined at any scale! 
In other words, the  real  $\Phi^3_6$-Kontsevich model  
could avoid triviality.

It is instructive to compare our exact results with a perturbative
BPHZ renormalisation of the model. In $D=6$ dimensions 
the full machinery of Zimmermann's forest formula 
\cite{Zimmermann:1969jj} is required. 
We provide in sec.~\ref{sec:BPHZ} the BPHZ-renormalisation of the
1-point function up to two-loop order. One of the contributing graphs
has an overlapping divergence with already 6 different forests. Individual
graphs show the full number-theoretic richness of quantum field
theory: up to two loops we encounter logarithms, polylogarithms 
$\mathrm{Li}_2$ and $\zeta(2)=\frac{\pi^2}{6}$. The amplitudes of the
graphs perfectly sum up to the Taylor expansion of the exact result.

The original BPHZ scheme
with normalisation conditions at a single scale leads in
just-renormalisable models to the renormalon problem which prevents
Borel resummation of the perturbation series. We  
also provide in sec.~\ref{sec:BPHZ} an example of a graph which 
shows the renormalon problem. But all these problems cancel
as our exact correlation functions are analytic(!) in the coupling
constant. Exact BPHZ renormalisation is fully consistent (for the
model under consideration)!

The most significant result of this paper is derived in
sec.~\ref{sec:positivity}. Matrix models such as the Kontsevich model
$\Phi^3_D$ arise from QFT-models on noncommutative geometry. The
prominent Moyal space gives rise to an external matrix $E$ having
linearly spaced eigenvalues with multiplicity reflecting the dimension
$D$. In \cite{Grosse:2013iva} two of us (H.G.+R.W.) have shown that
translating the type of scaling limit considered for the matrix model
correlation functions back to the position space formulation of the
Moyal algebra leads to Schwinger functions of an ordinary quantum
field theory on $\mathbb{R}^D$. Euclidean symmetry and invariance
under permutations are automatic. The most decisive
Osterwalder-Schrader axiom \cite{Osterwalder:1973dx,
  Osterwalder:1974tc}, \emph{reflection positivity}, amounts for the
Schwinger 2-point function to the verification that the diagonal
matrix model 2-point function is a Stieltjes function. We proved in
\cite{Grosse:2016pob} that for the $D=2$-dimensional Kontsevich model
this is \emph{not} the case. To our big surprise and exaltation, we
are able to prove:
\begin{thm}
The Schwinger 2-point function resulting from the scaling limit of the 
$\Phi^3_D$-QFT model on Moyal space with real coupling constant satisfies
reflection positivity in $D=4$ and $D=6$ dimensions. As such it is the 
Laplace-Fourier transform of the Wightman 2-point function
\begin{align}
\hat{W}_2(p_0,p_1,\dots,p_{D-1})= \frac{\theta(p_0)}{(2\pi)^{D-1}}  
\int_0^\infty dM^2 \,\varrho\Big(\frac{M^2}{\mu^2}\Big) 
\delta(p_0^2 -p_1^2-\dots-p_{D-1}^2-M^2) 
\end{align}
of a true
relativistic quantum field theory \cite{Streater:1964??} 
($\theta,\delta$ are the Heaviside
and Dirac distributions). Its K\"all\'en-Lehmann mass
spectrum $\varrho(\frac{M^2}{\mu^2})$ 
\cite{Kallen:1952zz, Lehmann:1954xi} is explicitly known and has
support on a scattering part 
with $M^2\geq 2\mu^2$ and an isolated fuzzy mass shell 
around $M^2=\mu^2$ of non-zero width. 
\end{thm}
The original Kontsevich model with purely imaginary coupling constant
cannot be positive. This property is shared with the $\lambda\Phi^4_4$
matricial quantum field theory \cite{Grosse:2012uv} where the stable
phase with $\lambda>0$, where the partition functions has a chance to
exist, cannot be reflection positive \cite{Grosse:2013iva}. For
$\lambda<0$, where the partition function is meaningless, a lot of
evidence has been given \cite{Grosse:2014lxa, Grosse:2015fka} for
reflection positivity of the 2-point function.

The exciting question whether also the higher
Schwinger functions of $\Phi^3_D$ are reflection positive 
in $D\in \{4,6\}$ is left for future investigation.

\bigskip

We dedicate this paper to the memory of Wolfhart Zimmermann. The
scientific community owes him, amongst others, the LSZ formalism,
momentum space Taylor subtraction under the integral, the forest
formula to handle overlapping divergences, operator product expansion
and the reduction-of-couplings scenario. Our
contribution applies several of these concepts to a non-perturbative 
setting. We are convinced that Wolfhart Zimmermann would have 
enjoyed these results.

One of us (H.G.) was several times invited (mostly by Julius Wess) to
the MPI in Munich at F\"ohringer Ring. But during these visits and
after seminars we got little by little good contacts to Wolfhart,
too. It was an honour to give a summary \cite{Grosse:2008zza} of our
earlier work with R.W.\ at the Ringberg Meeting celebrating the 80th
birthday of Wolfhart Zimmermann. And afterwards I enjoyed phone calls
from him showing his interests in our work.  We will remember Wolfhart
Zimmermann as a deep thinker and very kind person.

\section{The setup}

Let the `field' $\Phi=\Phi^*$ be a self-adjoint operator,
of finite rank $\leq \mathcal{N}$, on some
infinite-dimensional Hilbert space $\mathcal{H}$. 
In the end we are interested in a limit $\mathcal{N}\to \infty$ to
compact operators $\Phi$. Let $E$ be an unbounded self-adjoint positive
operator on $\mathcal{H}$ with compact resolvent.
We consider the following action functional
\begin{align}
S[\Phi]:=V \,\mathrm{tr}
\Big(ZE\Phi^2+ (\kappa+\nu E+\zeta E^2) \Phi +\frac{\lambda_{bare}
Z^{\frac{3}{2}}}{3}\Phi^3\Big),
\label{action-general}
\end{align}
where products and trace are understood in the algebra of finite rank
operators and only the projection $PEP$ to the finite-dimensional space
$(\ker \Phi)^\perp=P\mathcal{H}$ contributes to
(\ref{action-general}). The parameter $V$ is a constant discussed later,
  $\lambda_{bare}$ is the bare coupling
constant (real or complex) and $\kappa,\nu,\zeta,Z$ as well as
$\lambda_{bare}$ and the lowest eigenvalue $\mu_{bare}^2$ of $2E$
are functions of
$(V,\mathcal{N})$ and renormalised parameters $(\lambda_r,\mu^2)$.

The operator $E$ plays the r\^ole of a generalised Laplacian for which
the Theorem of H.\ Weyl applies on the recovery of the dimension from
the asymptotics of the spectrum:
\begin{defn}
The operator $E$ in (\ref{action-general})
encodes a \emph{spectral dimension}
\begin{align}
D =\inf\big\{p \in\mathbb{R}_+\;:~
\mathrm{tr}((1+E)^{-\frac{p}{2}}) < \infty\big\}.
\end{align}
\end{defn}
The dimension $D$ need not be an integer. We shall see, however, that
renormalisation is only sensitive to the even integer
$2[\frac{D}{2}]$.
\begin{itemize} \itemsep 0pt
\item In \cite{Grosse:2016pob} we treated $2[\frac{D}{2}]=2$
where only the renormalisation
parameter $\kappa$ 
is necessary: the other parameters are
$\nu=\zeta=0$, $Z=1$, $\mu_{bare}^2=\mu^2$, $\lambda_{bare}=\lambda$.

\item The
most involved case is $2[\frac{D}{2}]=6$, six dimensions in short. We
need all six renormalisation parameters
$\kappa,\nu,\zeta,Z,\lambda_{bare},\mu_{bare}^2$.

\item $2[\frac{D}{2}]=4$, four dimensions in short, is an
  intermediate case where we need
$\kappa,\nu,\mu_{bare}^2$ whereas $\zeta=0$, $Z=1$ and
$\lambda_{bare}=\lambda$.

\item $2[\frac{D}{2}] >6$ breaks the structures and cannot be renormalised.
\end{itemize}

Local details about the eigenvalues of $E$ (e.g.\ their degeneracy)
are not important. In the simplest case all eigenvalues are
different; here we would notationally proceed as
in \cite{Grosse:2016pob}. Spaces
with larger symmetry, such as spheres and tori, have degenerate
eigenvalues. In this paper we notationally assume a $\frac{D}{2}$-fold
cartesian product of a two-dimensional space with simple eigenvalues
$\{E_{|\under{n}|}\}_{|\under{n}|=0}^\infty$, growing at most linearly
  in $|\under{n}|$, which on the $D$-dimensional space arise with
multiplicity
$\binom{|\under{n}|+\frac{D}{2}-1}{\frac{D}{2}-1}$. This allows us
to label the eigenspaces by tuples
$\under{n}:=(n_1,\dots,n_{\frac{D}{2}})$, $n_i\in \mathbb{N}$,
of natural numbers (which include $0$). Indeed, defining
$|\under{n}|:=n_1+\dots +n_{\frac{D}{2}}$, there are
precisely $\binom{|\under{n}|+\frac{D}{2}-1}{\frac{D}{2}-1}$
different tuples
$\under{n}$ with the same $|\under{n}|$.
But again, these choices\footnote{induced by the natural matrix
  representation of noncommutative Moyal space} only affect the
notation in the first part;
any $E$ of the same $2[\frac{D}{2}]\leq 6$ leads to
a solvable model with the same renormalisation prescription
in the large-matrix limit.

From now on we assume that $D \in \{2,4,6\}$, that the projector $P$
commutes with $E$ and that the eigenvalues of $E$ are
a discretisation of a monotonously increasing $C^1$-function $e$,
\begin{align}
E=(E_{\under{n}} \delta_{\under{m},\under{n}}),\qquad
E_{\under{n}}:=\frac{\mu_{bare}^2}{2} + \mu^2
e\Big(\frac{|\under{n}|}{\mu^2 V^{\frac{2}{D}}}\Big),\qquad e(0)\equiv 0.
\label{En}
\end{align}
The parameter $\mu>0$
will become the renormalised mass, whereas the bare mass
$\mu_{bare}$ is a function of $(V,\mathcal{N},\lambda_r,\mu)$ identified later.
Note that $e$ being increasing amounts to a particular choice of the
projection $P$: Representing
$\Phi=(\Phi_{\under{m}\under{n}})_{\under{m},\under{n}\in
  \mathbb{N}^2}$ in this special eigenbasis, the rank condition becomes
$\Phi_{\under{m}\under{n}}=0$ if
$|\under{m}|>\mathcal{N}$ or
$|\under{n}|>\mathcal{N}$. In the very end we are interested in a
special limit $\mathcal{N}\to \infty$ which is
independent of the ordering.

In these conventions, and with
$\mathbb{N}^{D/2}_{\mathcal{N}}:=\{\under{m}\in \mathbb{N}^{\frac{D}{2}}\;:~
|\under{m}|\leq \mathcal{N}\}$, the action
(\ref{action-general}) takes the following form (simplifications
arise for $D=2$ and $D=4$):
\begin{align}
S[\Phi]&= V \Big(\sum_{\under{n},\under{m}\in \mathbb{N}^{D/2}_{\mathcal{N}}}
\!\!\! Z \Phi_{\under{m}\under{n}} \Phi_{\under{n}\under{m}}
\frac{H_{\under{m}\under{n}} }{2}
+ \!\! \sum_{\under{n}\in \mathbb{N}^{D/2}_{\mathcal{N}}}\!\!\!
(\kappa {+} \nu E_{\under{n}}{+}\zeta E_{\under{n}}^2 ) \Phi_{\under{n}\under{n}} +
\frac{\lambda_{bare} Z^{\frac{3}{2}}}{3} \!\!\!\!\!\!\!
\sum_{\under{n},\under{m},\under{l}\in \mathbb{N}^{D/2}_{\mathcal{N}}} \!\!\!\!\!
\Phi_{\under{n}\under{m}} \Phi_{\under{m}\under{l}}
\Phi_{\under{l}\under{n}} \Big),
\nonumber
\\
H_{\under{m}\under{n}} &:=
E_{\under{m}}+E_{\under{n}} .\label{Hmn}
\end{align}

The action formally defines a partition function
with external field $J$ (which is
also a self-adjoint matrix):
\begin{align}
\mathcal{Z}[J] &:= \int {\cal D}\Phi\; \exp \big( -S[\Phi]
+V\, {\rm tr} (J \Phi) \big)
\nonumber
\\
&= K \exp \Big( -\frac{\lambda_{bare} Z^{\frac{3}{2}}}{3V^2}
\sum_{\under{m},\under{n},\under{k} \in \mathbb{N}^{D/2}_{\mathcal{N}}}
\frac{\partial^3}{\partial J_{\under{m}\under{n}}
\partial J_{\under{n}\under{k}}
\partial J_{\under{k}\under{m}}}
\Big)
\mathcal{Z}_{free}[J],
\label{z_1}
\\
\mathcal{Z}_{free}[J] &:= \exp \Big(
\frac{V}{2} \!\!
\sum_{\under{m},\under{n} \in \mathbb{N}^{D/2}_{\mathcal{N}}}  \!\!\!\!
\frac{\big(J_{\under{m}\under{n}}
-(\kappa{+}\nu E_{\under{m}}{+}\zeta E_{\under{m}}^2)
\delta_{\under{m},\under{n}}\big)
\big(J_{\under{n}\under{m}}
-(\kappa{+}\nu E_{\under{m}}{+}\zeta E_{\under{m}}^2)
\delta_{\under{n},\under{m}}\big) }{Z H_{\under{n}\under{m}}}
\Big),
\label{z_free}
\end{align}
where $
K= \int {\cal D}\Phi \;\exp \big(-\frac{V Z}{2}
\sum_{\under{m},\under{n} \in \mathbb{N}^{D/2}_{\mathcal{N}}}
\Phi_{\under{m}\under{n}} H_{\under{m}\under{n}} \Phi_{\under{m}\under{n}} \big)$.

The integral in the first line of (\ref{z_1}) does
not make any sense for real $\lambda$. We rather view the partition
function as a tool to derive identities between the (formal) expansion
coefficients
$G_{|\under{p}_1^1\dots \under{p}_{N_1}^1|\dots|
\under{p}_1^B\dots \under{p}^B_{N_B}|}$ of the logarithm
\begin{align}
\log\frac{ \mathcal{Z}[J]}{\mathcal{Z}[0]}
=:\sum_{B=1}^\infty \; \sum_{1\leq N_1 \leq \dots \leq
  N_B}^\infty
\sum_{\under{p}_1^1,\dots,\under{p}^B_{N_B} \in \mathbb{N}^{D/2}_\mathcal{N}} \!\!\!\!
V^{2-B}
&\frac{G_{|\under{p}_1^1\dots \under{p}_{N_1}^1|\dots|\under{p}_1^B\dots \under{p}^B_{N_B}|}
}{S_{(N_1,\dots ,N_B)}}
\prod_{\beta=1}^B \frac{\mathbb{J}_{\under{p}_1^\beta\dots
    \under{p}^\beta_{N_\beta}}}{N_\beta},
\label{logZ}
\end{align}
where $\mathbb{J}_{\under{p}_1\dots
  \under{p}_{N_\beta}}:=\prod_{j=1}^{N_\beta} J_{\under{p}_j \under{p}_{j+1}}$, with
$N_\beta+1\equiv 1$, and with symmetry factor 
$S_{(N_1,\dots ,N_B)}=\prod_{i=1}^{s} \nu_i!$
if $(N_1,\dots,N_B)=(\underbrace{N'_1,\dots,N'_1}_{\nu_1},\dots,
\underbrace{N'_s,\dots,N'_s}_{\nu_s})$ for pairwise different $N_i'$.

These $(N_1{+}\dots{+}N_B)$-point functions
$G_{|\under{p}_1^1\dots \under{p}_{N_1}^1|\dots|
\under{p}_1^B\dots \under{p}^B_{N_B}|}$ are obtained by partial
$J$-derivatives of $\log\mathcal{Z}$ at $J=0$.
These external derivatives combine with the internal derivatives in
(\ref{z_1}) to identities between the $(N_1{+}\dots{+}N_B)$-point
functions called \emph{Schwinger-Dyson equations}. Our point of view is to
declare these Schwinger-Dyson equations (although their derivation was
purely formal) as \emph{exact identities between their rigorous
solutions}
$G_{|\under{p}_1^1\dots \under{p}_{N_1}^1|\dots|
\under{p}_1^B\dots \under{p}^B_{N_B}|}$. These Schwinger-Dyson
equations alone are not enough. As decisive tool we also need
\emph{Ward-Takahashi identities} which reflect the fact that the
partition function is unchanged under renaming the dummy integration
variable $\Phi$. This was first pointed out and used in
\cite{Disertori:2006nq}.
For actions $S[\Phi]=V\,\mathrm{Tr}(Z E\Phi^2 + \nu
E\Phi+\zeta E^2\Phi)+ S_{inv}[\Phi]$, where $S_{inv}$ is invariant under unitary
transformation $\Phi
\mapsto U^*\Phi U$, one has
\begin{align}
0 &= \int \mathcal{D}\Phi\;V\Big(Z(E\Phi^2-\Phi^2 E) +\nu (E\Phi-\Phi E)
+\zeta (E^2\Phi-\Phi E^2)
-(J\Phi+\Phi J)\Big)
\nonumber
\\
&\qquad\qquad\qquad\qquad\qquad \times
\exp\big(-S[\Phi]+V\,\mathrm{tr}(\Phi J)\big).
\label{Ward-E}
\end{align}
The $(\under{a}\under{b})$-component reads for diagonal $E$
\begin{align}
\frac{Z(E_{\under{a}}-E_{\under{b}})}{V}&
\sum_{\under{n}\in \mathbb{N}^{D/2}_{\mathcal{N}}}\frac{\partial^2 \mathcal{Z}[J]}{
\partial J_{\under{b}\under{n}}\partial J_{\under{n}\under{a}}}
+\nu
(E_{\under{a}}-E_{\under{b}}) \frac{\partial\mathcal{Z}[J]}{\partial
  J_{\under{b}\under{a}}}
+\zeta
(E_{\under{a}}^2-E_{\under{b}}^2) \frac{\partial\mathcal{Z}[J]}{\partial
  J_{\under{b}\under{a}}}
\nonumber
\\*
&= \sum_{\under{n}\in \mathbb{N}^{D/2}_{\mathcal{N}}}
\Big(J_{\under{a}\under{n}}\frac{\partial \mathcal{Z}[J]}{
\partial J_{\under{b}\under{n}}}-
J_{\under{n}\under{b}}\frac{\partial \mathcal{Z}[J]}{
\partial J_{\under{n}\under{a}}}
\Big).
\end{align}
We need a formula for
$\sum_{\under{n}\in \mathbb{N}^{D/2}_{\mathcal{N}}}\frac{\partial^2 \mathcal{Z}[J]}{
  \partial J_{\under{b}\under{n}}\partial J_{\under{n}\under{a}}}$,
na\"{\i}vely obtained by multiplication with
$\frac{V}{Z(E_{\under{a}}-E_{\under{b}})}$. But this needs
discussion. There is first a kernel $\mathrm{W}_{\under{a}}
\delta_{\under{a},\under{b}}$ in $\sum_{\under{n}\in
  \mathbb{N}^{D/2}_{\mathcal{N}}}\frac{\partial^2 \mathcal{Z}[J]}{
  \partial J_{\under{b}\under{n}}\partial J_{\under{n}\under{a}}}$
which was identified in \cite{Grosse:2012uv}. In the present case we
can restrict ourselves to $\under{a}\neq \under{b}$ where this kernel
is absent.  Remains the degeneracy problem: There is
$E_{\under{a}}-E_{\under{b}}=0$ even for $\under{a}\neq
\under{b}$. The solution, already sketched in \cite{Grosse:2012uv},
goes as follows: Because their definition via the action
(\ref{action-general}) and the definition (\ref{En}) of $E$
only involves the spectrum
of $E$, the $(N_1{+}\dots{+}N_B)$-point functions
$G_{|\under{p}_1^1\dots \under{p}_{N_1}^1|\dots| \under{p}_1^B\dots
  \under{p}^B_{N_B}|}$ only depend on $|\under{p}_i^\beta|$ (and
$\lambda$), but not on $\under{p}_i^\beta$ individually. If
we can afford to exclude the diagonal $\under{a} =\under{b}$, then we
can also afford to exclude $|\under{a}| =|\under{b}|$ and use the
following identity, valid for $|\under{a}|\neq |\under{b}|$,
\begin{align} \label{WT}
\sum_{\under{n}\in \mathbb{N}^{D/2}_{\mathcal{N}}}
\frac{\partial^2 \mathcal{Z}[J] }{\partial J_{\under{b}\under{n}}
\partial J_{\under{n}\under{a}} }
&=\sum_{\under{n}\in \mathbb{N}^{D/2}_{\mathcal{N}}}
\frac{V}{Z(E_{\under{a}}-E_{\under{b}}) }
\left(J_{\under{a}\under{n}}
\frac{\partial}{\partial J_{\under{b}\under{n}} }
-  J_{\under{n}\under{b}}
\frac{\partial}{\partial J_{\under{n}\under{a}} } \right) \mathcal{Z}[J]
-\frac{V}{Z}(\nu+\zeta H_{\under{a}\under{b}})
\frac{\partial \mathcal{Z}[J]}{\partial J_{|\under{b}\under{a}|}}.
\end{align}
Recall \cite{Grosse:2016pob} that for
$D=2$ we had $Z=1$ and $\nu=\zeta=0$, whereas
$Z=1$ and $\zeta=0$ for $D=4$. These are minor differences:
Only the prefactors of terms already present in the equations receive a
modification; no new structure is created. This is the essence of
multiplicative renormalisation. We will see that the new parameters
permit to regularise the integrals up to $D=6$. By extrapolation we would
need for $D\geq 8$ a term $\propto E^3\Phi$ in the action. But such a
term gives rise to
$E_{\under{a}}^2+E_{\under{a}}E_{\under{b}}+E_{\under{b}}^2$ in
(\ref{WT}) which completely destroys the previous structures. Hence,
renormalisability is lost for $D\geq 8$.

\section{Schwinger-Dyson equations and 
solution for $B=1$}

\subsection{Equations}

In the same way as in our previous paper \cite{Grosse:2016pob}, namely
by inserting (\ref{z_1}), (\ref{z_free}) into the corresponding term of
(\ref{logZ}),  we
derive formulae for the connected $N$-point functions. We have to
discuss separately the cases $N=1$, $N=2$ and $N\geq 3$.
The equation for
$G_{|\under{a}|}$ is straightforward:
\begin{align}
G_{|\under{a}|}&=\frac{1}{Z H_{\under{a}\under{a}}}
\Big\{-\kappa-\nu  E_{\under{a}}-\zeta E_{\under{a}}^2
-\lambda_{bare} Z^{\frac{3}{2}} \Big( 
G_{|\under{a}|}^2
+\frac{1}{V} \!\!\! \sum_{\under{m}\in
\mathbb{N}^{D/2}_{\mathcal{N}}} \!\!
G_{|\under{a}\under{m}|} +\frac{G_{|\under{a}|\under{a}|}}{V^2} 
\Big)\Big\}.
\label{1point_ward}
\end{align}
The equation for
$G_{|\under{a}\under{b}|}$ with
$|\under{a}|\neq |\under{b}|$ reads:
\begin{align}
G_{|\under{a}\under{b}|}
&=\frac{1}{Z H_{\under{a}\under{b}}}-\frac{\lambda_{bare} Z^{\frac{1}{2}}}{V^2
H_{\under{a}\under{b}}} \frac{1}{\mathcal{Z}[0]}
\sum_{\under{m}\in
\mathbb{N}^{D/2}_{\mathcal{N}}}
\frac{\partial}{\partial J_{\under{a}\under{b}}}
\frac{\partial}{\partial J_{\under{b}\under{m}}}
\frac{\partial}{\partial J_{\under{m}\under{a}}}
\mathcal{Z}[J]
\Big|_{J=0}
\nonumber
\\
&=\frac{1}{ZH_{\under{a}\under{b}}}\Big(1+\lambda_{bare} Z^{\frac{1}{2}}
\frac{(G_{|\under{a}|}-G_{|\under{b}|})}{E_{\under{a}}-E_{\under{b}}}
+ \lambda_{bare} Z^{\frac{1}{2}} (\nu +\zeta H_{\under{a}\under{b}})
G_{|\under{a}\under{b}|}
\Big)\;.
\label{2pointWard}
\end{align}
The last line is obtained by inserting the Ward-Takahashi
identity (\ref{WT}) and comparing the resulting $J$-derivatives with
(\ref{logZ}).
For $(N{>}2)$ and pairwise different
$|\under{a}_i|$ we have
\begin{align}
&G_{|\under{a}_1  \under{a}_2  \dots  \under{a}_N|}
\label{GN-a}
\\
&=\frac{(-\lambda_{bare}) Z^{\frac{1}{2}}}{V^2 H_{\under{a}_1\under{a}_2} \mathcal{Z}[0]}
\sum_{\under{n}\in \mathbb{N}^{D/2}_{\mathcal{N}}}
\frac{\partial}{\partial J_{\under{a}_2 \under{a}_3}} \cdots
\frac{\partial}{\partial J_{\under{a}_{N}\under{a}_1}}
\frac{\partial}{\partial J_{\under{a}_1\under{n}}}
\frac{\partial}{\partial J_{\under{n}\under{a}_2}}
\mathcal{Z}[J]
\Big|_{J=0}
\nonumber
\\
&=\frac{(-\lambda_{bare}) Z^{\frac{1}{2}}}{V H_{\under{a}_1\under{a}_2}
  \mathcal{Z}[0]}
\frac{\partial}{\partial J_{\under{a}_2 \under{a}_3}} \cdots
\frac{\partial}{\partial J_{\under{a}_{N}\under{a}_1}}
\Big(
\!\sum_{\under{n}\in \mathbb{N}^{D/2}_{\mathcal{N}}}\!
\frac{\big(J_{\under{a}_2\under{n}}
\frac{\partial \mathcal{Z}[J]  }{\partial J_{\under{a}_1\under{n}}}
-J_{\under{n}\under{a}_1}
\frac{\partial \mathcal{Z}[J]  }{\partial J_{\under{n}\under{a}_2}}
\big)}{Z(E_{\under{a}_2}-E_{\under{a}_1})}
-\frac{\nu {+}\zeta H_{\under{a}_1\under{a}_2}}{Z}
\frac{\partial \mathcal{Z}[J]  }{\partial J_{\under{a}_1\under{a}_2}}
\Big)\Big|_{J=0}.
\nonumber
\end{align}
The step to the last line uses the Ward-Takahashi identity (\ref{WT})
for pairwise different 1-norms of indices.

We start to discuss (\ref{GN-a}) because this unambiguously fixes
a particular combination of
the renormalisation parameters.
We multiply (\ref{GN-a})  by
$H_{\under{a}_1\under{a}_2}$, perform the $J$-differentiations and
collect all coefficients of
$G_{|\under{a}_1  \under{a}_2  \dots  \under{a}_N|}$ on the lhs:
\begin{align}
\big(1{-}Z^{-\frac{1}{2}}\lambda_{bare} \zeta\big)
\Big(E_{\under{a}_1}{+}E_{\under{a}_2}
-\frac{ Z^{-\frac{1}{2}}\lambda_{bare}\nu}{1{-}Z^{-\frac{1}{2}}\lambda_{bare} \zeta}\Big)
G_{|\under{a}_1  \under{a}_2  \dots  \under{a}_N|}
&=\lambda_{bare} Z^{\frac{1}{2}}
\frac{G_{|\under{a}_1\under{a}_3\dots \under{a}_{N}|}
-G_{|\under{a}_2\under{a}_3\dots \under{a}_{N}|}}{Z(E_{\under{a}_1}
-E_{\under{a}_2})} .
\end{align}
In terms of the functions
\begin{align}
F_{\under{a}}=E_{\under{a}}
-\frac{ Z^{-\frac{1}{2}}\lambda_{bare}\nu}{2(1-Z^{-\frac{1}{2}}\lambda_{bare}
  \zeta)},\qquad
\lambda_r:= \frac{\lambda_{bare}}{Z^{\frac{1}{2}}
\big(1-Z^{-\frac{1}{2}}\lambda_{bare} \zeta\big) },
\label{def:F}
\end{align}
\emph{which necessarily must be finite}, we thus obtain
\begin{align}
G_{|\under{a}_1  \under{a}_2  \dots  \under{a}_N|}
=\lambda_r
\frac{G_{|\under{a}_1\under{a}_3\dots \under{a}_{N}|}
-G_{|\under{a}_2\under{a}_3\dots \under{a}_{N}|}}{(F_{\under{a}_1}^2
-F_{\under{a}_2}^2)}.
\label{NptWT}
\end{align}
The same steps yield for (\ref{2pointWard}):
\begin{align}
G_{|\under{a}\under{b}|}
&=\frac{1}{Z(1-Z^{-\frac{1}{2}}\lambda_{bare} \zeta)
(F_{\under{a}}+F_{\under{b}} )}
+\lambda_r
\frac{G_{|\under{a}|}-G_{|\under{b}|}}{F_{\under{a}}^2-F_{\under{b}}^2}.
\label{2pointWard-a}
\end{align}

Now we describe the renormalisation.  We clearly need well-defined
functions $G_{|\under{a}|},F_{\under{a}},\lambda_r$ in the limit
$\mathcal{N}\to \infty$.  This in turn forces a particular singular
behaviour of $\kappa,\nu,\zeta,Z,\mu_{bare},\lambda_{bare}$.  As
familiar from perturbative renormalisation, there is still
considerable freedom in choosing finite terms of these singular
functions. We follow Zimmermann's prescription \cite{Zimmermann:1969jj}
and fix the finite terms by 
normalisation of the first Taylor expansion coefficients of 
relevant and marginal correlation functions
(with $i=1,\dots,\frac{D}{2}$ below):
\begin{align}
D\geq 2:&&&G_{|\under{0}|}=0,\qquad
\label{normalisation}
\\
D\geq 4:&&&
\frac{\partial}{\partial a_i} G_{|\under{a}|}\Big|_{\under{a}=0}=0,\qquad
G_{|\under{0}\under{0}|}=\frac{1}{\mu^2},\qquad
\nonumber
\\
D= 6:&&&
\frac{\partial^2}{\partial a_i^2} G_{|\under{a}|}\Big|_{\under{a}=0}=0,\qquad
\frac{\partial}{\partial a_i} G_{|\under{a}\under{b}|}
\Big|_{\under{a}=\under{b}=0}=
\frac{\partial}{\partial b_i} G_{|\under{a}\under{b}|}
\Big|_{\under{a}=\under{b}=0}=
-\frac{1}{\mu^4 V^{\frac{2}{D}}} e'(0).
\nonumber
\end{align}
Inserting the conditions on $G_{|\under{a}|}$
into (\ref{2pointWard-a}) reduces the conditions on
$G_{|\under{a}\under{b}|}$ for $D\geq 4$ to
\[
\frac{1}{Z(1{-}Z^{-\frac{1}{2}}\lambda_{bare} \zeta)
(2F_{\under{0}})}=\frac{1}{\mu^2},
\qquad
-\frac{1}{Z(1{-}Z^{-\frac{1}{2}}\lambda_{bare} \zeta)
(2F_{\under{0}})^2} \frac{e'(0)}{V^{\frac{2}{D}}}
=-\frac{1}{\mu^4 V^{\frac{2}{D}}} e'(0)
\]
with solution
\begin{align}
Z(1-Z^{-\frac{1}{2}}\lambda_{bare} \zeta)=1,\qquad
F_{\under{0}}=\frac{\mu_{bare}^2}{2}
-\frac{ Z^{-\frac{1}{2}}\lambda_{bare}\nu}{2(1-Z^{-\frac{1}{2}}\lambda_{bare}
  \zeta)}
=
\frac{\mu^2}{2}.
\label{F0}
\end{align}
We thus conclude in (\ref{def:F})
\begin{align}
\lambda_r:=Z^{\frac{1}{2}}\lambda_{bare},\qquad
E_{\under{a}}=F_{\under{a}}+\frac{1}{2}\lambda_r\nu,\qquad
\frac{\lambda_r\zeta}{Z}=1-\frac{1}{Z}
\label{EF}
\end{align}
and define
\begin{align}
\frac{W_{|\under{a}|}}{2\lambda_r} := G_{|\under{a}|}+
\frac{F_{\under{a}}}{\lambda_r} \qquad \Rightarrow\qquad
G_{|\under{a}\under{b}|}= \frac{1}{2}
\frac{W_{|\under{a}|}-W_{|\under{b}|}}{
F_{\under{a}}^2-F_{\under{b}}^2}.
\label{2pointWard-b}
\end{align}

The above identities and definitions are now inserted into
(\ref{1point_ward}). To make sense out of 
$\sum_{\under{m}\in
\mathbb{N}^{D/2}_{\mathcal{N}}} 
G_{|\under{a}\under{m}|}$ it is necessary 
to permit again the case
$|\under{a}|=|\under{b}|$ in (\ref{2pointWard-a}).
This is achieved by a continuity argument\footnote{Alternatively, 
the limit $V\to \infty$ permits a cheaper solution: We can restrict
the sum to
$\sum_{\under{m}\in
\mathbb{N}^{D/2}_{\mathcal{N}},|\under{m}|\neq |\under{a}|} 
G_{|\under{a}\under{m}|}$ because 
$\frac{1}{V}\sum_{|\under{m}|=|\under{a}|}
G_{|\under{a}\under{m}|}\to 0$. For the sake of easier notation we 
write unrestricted sums. We leave it to the reader's taste 
to exclude
the terms ``$\frac{0}{0}$'' or to define them by continuity.\label{fn2}},
$G_{|\under{a}\under{a}|}=\frac{1}{2F_{\under{a}}}
\big(1+\lambda_r \lim_{|\under{b}|\to
  |\under{a}|}\frac{(G_{|\under{a}|}-G_{|\under{b}|})}{F_{\under{a}}-F_{\under{b}}}\big)$.
The limit is well-defined in
perturbation theory where $G_{|\under{a}|}$ is, before performing the loop
sum, a rational function of the $\{F_{\under{n}}\}$ so that a
factor $F_{\under{a}}-F_{\under{b}}$ can
be taken out of $G_{|\under{a}|}-G_{|\under{b}|}$. The result is
\begin{align}
W_{|\under{a}|}^2
+ 2\lambda_r \nu W_{|\under{a}|}&=
\frac{4}{Z} F_{\under{a}}^2
-\Big(4\frac{\lambda_r \kappa}{Z}
+\Big(1{+}\frac{1}{Z}\Big) (\lambda_r \nu)^2\Big)
-\frac{2\lambda_r^2 }{V} \!\!\!\sum_{\under{n}\in
\mathbb{N}^{D/2}_{\mathcal{N}}} \!\!\!
\frac{W_{|\under{a}|}-W_{|\under{n}|}}{
F_{\under{a}}^2-F_{\under{n}}^2}
-\frac{4\lambda_r^2}{V^2} G_{|\under{a}|\under{a}|}.
\label{1point_ward-b}
\end{align}

\subsection{Large-($\mathcal{N},V$) limit
and integral equations}

The action $S[\Phi]$, hence $\log \mathcal{Z}$ and
$G_{|\under{n}|},W_{|\under{n}|}$ only depend (via $E_{\under{n}}$
defined in (\ref{En}))
on the 1-norm $|\under{n}|=n_1+\dots+n_{D/2}$ and not individually on
the components $n_i$ of $\under{n}$.
Therefore, the sum in (\ref{1point_ward-b}) translates
into
\[
\sum_{\under{m}\in \mathbb{N}^{D/2}_{\mathcal{N}}} f(|\under{m|})=
\sum_{|\under{m}|=0}^{\mathcal{N}} \binom{|\under{m}|+\frac{D}{2}-1}{
\frac{D}{2}-1}
f(|\under{m|}),
\]
giving
\begin{align}
W_{|\under{a}|}^2
+2 \lambda_r \nu W_{|\under{a}|}
-\frac{4}{Z}F_{\under{a}}^2
+\frac{2\lambda_r^2}{V} \sum_{|\under{m}|=0}^{\mathcal{N}}
\binom{|\under{m}|+\frac{D}{2}-1}{
\frac{D}{2}-1}
\frac{W_{|\under{a}|}-W_{|\under{m}|}}{
F_{\under{a}}^2-F_{\under{m}}^2}
+\frac{4\lambda_r^2}{V^2}G_{|\under{a}|\under{a}|}
= \text{const}.
\label{1point_ward-c}
\end{align}
We take the limit $\mathcal{N},V \rightarrow \infty$ subject to
fixed ratio
\begin{align}
\frac{\mathcal{N}}{V^{\frac{2}{D}}}=\mu^2 \Lambda^2,
\end{align}
in which the sum converges to a Riemann integral
\begin{align}
\lim \frac{1}{V^{\frac{2}{D}}} \sum_{m=0}^{\mathcal{N}} f(m/V^{\frac{2}{D}})
= \mu^2\Lambda^2 \int_0^1 d\tau\;
f(\mu^2 \Lambda^2 \tau)
= \mu^2 \int_0^{\Lambda^2} dx\;f(\mu^2 x).
\end{align}
Expressing the 1-norms of discrete matrix elements as
$|\under{a}|=:V^{\frac{2}{D}}\mu^2 x$, with
$x\in [0,\Lambda^2]$, and taking the normalisation
$2F_{\under{0}}=\mu^2$ from (\ref{F0}) and the
relations (\ref{def:F})+(\ref{En}) into account, we arrive at
$F_{\under{a}}\big|_{|\under{a}|=:V^{\frac{2}{D}}\mu^2 x}
=\mu^2(e(x)+\frac{1}{2})$.

The mass $\mu$ is the only dimensionful parameter. The previously
introduced functions and parameters have in spectral dimension
$D\in \{2,4,6\}$ the following mass dimensions:
\begin{align}
&[E_{\under{a}}]=[F_{\under{a}}]=[W_{|\under{a}|}]=\mu^2,\quad
[V]=\mu^{-D},\quad [Z]=\mu^0,\quad
[\Phi]=\mu^{\frac{D-2}{2}},\quad
[J]=\mu^{\frac{D+2}{2}},
\nonumber
\\*
&[\lambda_{bare}]=[\lambda_{r}]= \mu^{3-\frac{D}{2}},\quad
[\kappa]=\mu^{\frac{D+2}{2}},\quad
[\nu]=\mu^{\frac{D-2}{2}},\quad
[\zeta]=\mu^{\frac{D-6}{2}},\quad
\nonumber
\\*
&[G_{|\under{p}^1_1\dots\under{p}^1_{N_1}|\dots|
\under{p}^B_1\dots\under{p}^B_{N_B}|}]=\mu^{(2-B-\frac{N}{2})D-N},
\end{align}
where $N=N_1+\dots+N_B$. In terms of the dimensionless functions
\begin{align}
&W_{|\under{a}|}\big|_{\under{a}=:V^{\frac{2}{D}}\mu^2 x}=:\mu^2
\tilde{W}(x),\qquad
\tilde{\lambda}:=\mu^{\frac{D}{2}-3}\lambda_r,\qquad
\tilde{\nu}:=\mu^{1-\frac{D}{2}} \nu,
\nonumber
\\
&G_{|\under{p}^1_1\dots\under{p}^1_{N_1}|\dots|
\under{p}^B_1\dots\under{p}^1_{N_B}|}\big|_{|\under{p}^\beta_i|
=V^{\frac{2}{D}}\mu^2 x^\beta_i}
=: \mu^{(2-B-\frac{N}{2})D-N}
\tilde{G}(x^1_1,\dots, x^1_{N_1}|\dots|x^B_1,\dots, x^1_{N_B}),
\end{align}
the large-$(\mathcal{N},V)$ limit of $\mu^{-4}$ times
(\ref{1point_ward-c}) takes the form
\begin{align}
&\tilde{W}^2(x)
+2 \tilde{\lambda} \tilde{\nu} \tilde{W}(x)-\frac{(2e(x){+}1)^2}{Z}
+\frac{2 \tilde{\lambda}^2}{(\frac{D}{2}{-}1)!}
\int_0^{\Lambda^2} \!\!\! dt\;
t^{\frac{D}{2}-1}
\frac{\tilde{W}(x)-\tilde{W}(t)}{
\big(e(x){+}\frac{1}{2}\big)^2 -\big(e(t){+}\frac{1}{2}\big)^2 }
=\text{const}.
\label{Int_eq_field0}
\end{align}
We have used here the fact proved later that
$G_{|\under{a}|\under{a}|}$ has a finite large-$(\mathcal{N},V)$ limit
so that $\frac{4\lambda_r^2}{V^2}G_{|\under{a}|\under{a}|}$ from
(\ref{1point_ward-c}) does not contribute to the limit.  A final
transformation
\begin{align}
X:= (2e(x)+1)^2 ,\quad
W(X)=\tilde{W}(x(X)),\quad
G(X)=\tilde{G}(x(X)),\quad
\end{align}
and similarly for other capital letters $Y(y),T(t)$ and functions
$G(X,Y)=\tilde{G}(x(X),y(Y))$ etc.,
simplifies (\ref{Int_eq_field0}) to
\begin{align}
(W(X))^2&+2\tilde{\lambda}\tilde{\nu} W(X)
+\int_1^{\Xi} dT \;\rho(T) \,
\frac{W(X)-W(T)}{X-T} -X
= \text{const},
\label{Int_eq_field1}
\\
\rho(T)&:=\frac{2\tilde{\lambda}^2\big(e^{-1}(\tfrac{\sqrt{T}-1}{2})
\big)^{\frac{D}{2}-1}}{(\frac{D}{2}{-}1)!\sqrt{T}
\cdot e'(e^{-1}(\frac{\sqrt{T}-1}{2}))}
,\qquad \Xi:=(1+2e(\Lambda^2))^2.
\nonumber
\end{align}
Building on \cite{Makeenko:1991ec} we proved in 
\cite[eq.\ (4.14)]{Grosse:2016pob} that (\ref{Int_eq_field1})
is solved by 
\begin{align}
W(X) &:= \frac{\sqrt{X+c}}{\sqrt{Z} }
-\tilde{\lambda}\tilde{\nu}
+ \frac{1}{2} \int_1^\Xi dT \frac{\rho(T)}{(\sqrt{X+c}
+ \sqrt{T+c})\sqrt{T+c}},
\label{MS-solution}
\end{align}
for some function $c(\tilde{\lambda},e,\tilde{\nu},Z)$. 
The functions 
$c,\tilde{\nu},Z$ are fixed by the normalisation conditions
(\ref{normalisation}) which translate into
$W(X)=\sqrt{X}+\mathcal{O}((\sqrt{X}-1)^{\frac{D}{2}})$, i.e.\ 
\begin{align}
\underbrace{W(1)=1}_{D\geq 2},\qquad \underbrace{
W'(1)=\frac{d}{dX} \sqrt{X}\Big|_{X=1}=\frac{1}{2}}_{D\geq 4}
 ,\qquad \underbrace{
W''(1)=\frac{d^2 }{dX^2} \sqrt{X}\Big|_{X=1}=-\frac{1}{4}}_{D=6}.
\label{normalisation-1}
\end{align}
The first condition fixes $\tilde{\nu}(c,Z)$ to
\begin{align}
1 &=: \frac{\sqrt{1+c}}{\sqrt{Z} }
-\tilde{\lambda}\tilde{\nu}
+ \frac{1}{2} \int_1^\Xi dT \frac{\rho(T)}{(\sqrt{1+c}
+ \sqrt{T+c})\sqrt{T+c}}.
\label{MS-solution-0}
\end{align}
For $D=2$ we had $Z=1$ and $\nu=\tilde{\nu}=0$ so that in the limit
$\Xi\to \infty$ (safe in $D=2$ where $\rho(T)\propto
\frac{1}{\sqrt{T}}$) we recover the solution of
\cite[eq.\ (4.15)]{Grosse:2016pob}. For $D\geq 4$ equations
(\ref{MS-solution})+(\ref{MS-solution-0}) read
\begin{align}
W(X) &= 1+(\sqrt{X{+}c}-\sqrt{1{+}c})
\Big(\frac{1}{\sqrt{Z}}
- \frac{1}{2} \int_1^\Xi \!\!\!
\frac{dT\; \rho(T)}{(\sqrt{X{+}c} + \sqrt{T{+}c})
(\sqrt{1{+}c} + \sqrt{T{+}c})\sqrt{T{+}c}}\Big).
\label{MS-solution-a}
\end{align}
The condition on $W'(1)$ in (\ref{normalisation-1}) then fixes $Z(c)$ to
\begin{align}
\frac{1}{2} &= \frac{1}{2\sqrt{1+c}}
\Big(\frac{1}{\sqrt{Z}}- \frac{1}{2} \int_1^\Xi dT
\frac{\rho(T)}{(\sqrt{1+c} + \sqrt{T+c})^2\sqrt{T+c}}\Big).
\label{MS-solution-b}
\end{align}
For $D=4$ where $Z\equiv 1$, this equation determines the main
function $c(\tilde{\lambda},e)$ (in the limit $\Xi \to \infty$ which
is safe for $\rho(T)\propto \text{const}$)
to
\begin{align}
D=4: && 1-\sqrt{1+c} &= \frac{1}{2} \int_1^\infty dT
\frac{\rho(T)}{(\sqrt{1+c} + \sqrt{T+c})^2\sqrt{T+c}} .
\label{MS-solution-b4}
\end{align}
The formula differs from $D=2$ by a power of 2 in
the denominator $(\sqrt{1+c} + \sqrt{T+c})$. With this solution
for $c$ we have
\begin{align}
D{=}4: && W(X) &= \sqrt{X{+}c}+1{-}\sqrt{1{+}c}
- \frac{1}{2} \int_1^\infty \!\!\!
\frac{dT\; \rho(T)\;(\sqrt{X{+}c}-\sqrt{1{+}c})
}{(\sqrt{X{+}c} + \sqrt{T{+}c})
(\sqrt{1{+}c} + \sqrt{T{+}c})\sqrt{T{+}c}}.
\label{MS-solution-a4}
\end{align}

For $D=6$ we conclude
\begin{align}
\frac{1}{\sqrt{Z}} =
\sqrt{1+c} + \frac{1}{2} \int_1^\Xi dT
\frac{\rho(T)}{(\sqrt{1+c} + \sqrt{T+c})^2\sqrt{T+c}}.
\label{1/sqrtZ}
\end{align}
This allows us to compute the $\beta$-function of the
running coupling constant\footnote{For the sake of readability
we assume here that $c$ is independent of $\Xi$ and given by
(\ref{MS-solution-d}) below. Requiring $W''(1)=-\frac{1}{4}$ exactly
restricts the integral in (\ref{MS-solution-d}) to $[1,\Xi]$ and leads
to $c(\tilde{\lambda},e,\Xi)$. We are free to admit a finite
renormalisation of $W''(1)$ which approaches $-\frac{1}{4}$ only in
the limit $\Xi\to \infty$.}
$\lambda_b(\Xi)= \frac{1}{\sqrt{Z(\Xi)}} \lambda_r$:
\begin{align}
\beta_\lambda &:= \Lambda^2\frac{d\lambda_{bare}(\Xi(\Lambda))}{d\Lambda^2}
= 4 \sqrt{\Xi} e^{-1}(\tfrac{\sqrt{\Xi}-1}{2})
\cdot e'(e^{-1}(\tfrac{\sqrt{\Xi}-1}{2}))
\frac{d}{d\Xi} \lambda_b(\Xi)
\nonumber
\\*
&= \frac{2\lambda_r^3  \Lambda^6}{
\big(\sqrt{1+c} + \sqrt{(2e(\Lambda^2)+1)^2+c}\big)^2
\sqrt{(2e(\Lambda^2)+1)^2+c}}>0.
\label{MS-solution-b6}
\end{align}
We learn that the $\beta$-function is -- for real coupling constant --
strictly positive\footnote{More precisely, $\beta_\lambda$ has the
  same sign as $\lambda$ which means that
  $|\lambda_{bare}(\Lambda^2)|$ increases
with $\Lambda^2$.}, with $\lambda_b(\Xi)\stackrel{\Xi\to
  \infty}{\longrightarrow} +\infty$, \emph{but without developing a Landau
pole} (a singularity of $\lambda_b$ already at finite
$\Xi_0$). We also have $Z(\Xi)\in [0,1]$, as it should.
We remark that \cite{Grosse:2006tc} addresses the case that
$\lambda_{bare},\lambda_r 
\in \mathrm{i}\mathbb{R}$, hence $\lambda_r^2<0$, resulting in negative
$\beta$-function and asymptotic freedom. We would like to point out,
however, that this also implies $\rho(T)<0$ and consequently 
the senseless result $\frac{1}{\sqrt{Z}} \to - \infty$. We shall see
in the final section that an important positivity property only holds
for $\lambda_r$ real. 

Inserted into (\ref{MS-solution-a}) we get
\begin{align}
D=6:&& W(X) &= \sqrt{X+c}\sqrt{1+c} -c
\nonumber
\\*
&&& +\frac{1}{2}
\int_1^\infty
\frac{dT\; \rho(T)\;(\sqrt{X+c}-\sqrt{1+c})^2}{(\sqrt{X+c} + \sqrt{T+c})
(\sqrt{1+c} + \sqrt{T+c})^2\sqrt{T+c}},
\label{MS-solution-c}
\end{align}
where the limit $\Xi\to \infty$ is now safe.
Eventually, the condition on $W''(1)$ in (\ref{normalisation-1})
determines $c(\tilde{\lambda},e)$ to
\begin{align}
-c &= \int_1^\infty
\frac{dT\; \rho(T)}{(\sqrt{1+c} + \sqrt{T+c})^3\sqrt{T+c}}.
\label{MS-solution-d}
\end{align}

Obviously, $\tilde{\lambda}=0$ and hence $\rho=0$
corresponds to $c=0$ for any $D$. For given $e(x)$,
thus $\rho(T)$, the
implicit function theorem provides a unique diffeomorphism
$\tilde{\lambda}^2\mapsto c(\tilde{\lambda},e)$ on a neighbourhood
of $0 \in \mathbb{R}$ or $0\in \mathbb{C}$. Since we will
be able to express all correlation functions in terms of elementary
functions of $c(\tilde{\lambda},e)$ and $\rho(\tilde{\lambda},e)$,
this proves analyticity of all correlation functions in these neighbourhoods.

According to
\cite[Prop 4.2]{Grosse:2016pob} (which is unchanged
due to (\ref{NptWT})),
all $N$-point functions are explicitly computable from this
solution $W(X)$:
\begin{prop} \label{prop_Npt_1}
The connected ($N{\geq} 1$)-point functions in the scaling limit
$\tilde{G}(x_1,\dots,x_N)=\lim_{V,\mathcal{N}\to \infty}
\mu^{N+(\frac{N}{2}-1)D}G_{|\under{p}_1\dots,\under{p}_N|}\big|_{
|\under{p}_i|=V^{2/D}  \mu^2x_i}$
and $G(X_1,\dots,X_n):=\tilde{G}(x_1(X_1),\dots,x_N(X_N))$ read
\begin{align}
G(X_1) &:=\frac{W(X_1)-\sqrt{X_1}}{2\tilde{\lambda}}\qquad \text{for }N=1,
\\
G(X_1,\dots,X_N)
&=
\sum_{k=1}^{N} \frac{W(X_k)}{2\tilde{\lambda}}
\prod_{l=1, l\neq k}^{N} \frac{4\tilde{\lambda}}{X_k-X_l}
\qquad \text{for }N>1,
\label{GXN}
\end{align}
where $W(X)$ is given in
\cite[eq.\ (4.13)+(4.15)]{Grosse:2016pob} for $D=2$, in
(\ref{MS-solution-b})+(\ref{MS-solution-a4}) for $D=4$ and in
(\ref{MS-solution-c})+(\ref{MS-solution-d}) for $D=6$.
We have $\tilde{\lambda}=\lambda_r$ for $D=6$ and
$\tilde{\lambda}=\frac{\lambda}{\mu^{3-D/2}}$ for $D\in \{2,4\}$.
\end{prop}


\section{$N$-points function with
$B \ge 2$ boundaries}

The derivation of the formula for the $(N_1{+}\dots{+}N_B$)-point function
with one $N_i>1$ proceeds along the same lines as in our previous paper
\cite{Grosse:2016pob}. Key ingredient is the
Ward-Takahashi identity (\ref{WT}) which compared with
\cite{Grosse:2016pob} contains additional parameters
$\nu,\zeta\neq 0$ and $Z\neq 1$. However, in identical manner as in
the derivation of (\ref{NptWT}), the resulting contributions
reconstruct the renormalised functions
$F_{\under{a}}$ instead of $E_a$ in \cite{Grosse:2016pob}. We thus have
for $N_1>1$
\begin{align}
G_{|\under{a}^1_1\dots \under{a}^1_{N_1}|\dots|\under{a}^B_1\dots \under{a}_{N_B}^B|}
= \lambda_r \frac{G_{|\under{a}^1_1\under{a}^1_3\dots \under{a}^1_{N_1}|\under{a}^2_1\dots \under{a}^2_{N_2}|\dots|\under{a}^B_1\dots \under{a}_{N_B}^B|}
-G_{|\under{a}_2^1\under{a}_3^1\dots \under{a}^1_{N_1}|\under{a}^2_1\dots \under{a}^2_{N_2}|\dots|\under{a}^B_1\dots \under{a}_{N_B}^B|}}{F_{\under{a}_1^1}^2-
F_{\under{a}_2^1}^2}.
\label{GNB}
\end{align}
Its reduction to $(1{+}\dots{+}1)$-point functions is,
up to renormalisations $E\mapsto F$ and $\lambda\mapsto \lambda_r$,
identically as in \cite[Prop. 5.2]{Grosse:2016pob}:
\begin{prop} \label{prop_Npt_B}
Let $B\geq 2$. The connected ($N_1{+}\dots{+}N_B$)-point function
with one $N_i>1$ is given for any $D\in \{2,4,6\}$ by
\begin{align}
&
G_{|\under{a}^1_1\dots \under{a}^1_{N_1}|\dots|\under{a}^B_1\dots \under{a}_{N_B}^B|}
\label{GNB-final}
\\
&=\lambda_r^{N_1+\dots+N_B-B}
\sum_{k_1=1}^{N_1} \dots
\sum_{k_B=1}^{N_B}
G_{|\under{a}^1_{k_1}|\dots|\under{a}^B_{k_B}|}
\Big(\prod_{l_1=1, l_1\neq k_1}^{N_1} \!\!\!
\frac{1}{F_{\under{a}^1_{k_1}}^2 {-}F_{\under{a}^1_{l_1}}^2}
\Big)\cdots
\Big(\prod_{l_B=1, l_B\neq k_B}^{N_B}\!\!\!
\frac{1}{F_{\under{a}^B_{k_B}}^2 {-}F_{\under{a}^B_{l_B}}^2}
\Big)
\nonumber
\end{align}
(where $\lambda_r\equiv \lambda$ for $D\in \{2,4\}$), its 
large-$(\mathcal{N},V)$ limit by
\begin{align}
&G(X^1_1,\dots, X^1_{N_1}|\dots|X^B_1,\dots, X_{N_B}^B)
\label{GNB-final-X}
\\
&=\tilde{\lambda}^{N_1+\dots+N_B-B}
\sum_{k_1=1}^{N_1} \dots
\sum_{k_B=1}^{N_B}
G(X^1_{k_1}|\dots|X^B_{k_B})
\prod_{\beta=1}^B \prod_{l_\beta=1, l_\beta\neq k_\beta}^{N_\beta}
\frac{4}{X^\beta_{k_\beta}-X^\beta_{l_\beta}}.
\nonumber
\end{align}
\end{prop}

\subsection{$(1{+}\dots{+}1)$-point function}

The Schwinger-Dyson equation for the
$(1{+}\dots{+}1)$-point function is at an intermediate step and up to
taking multiple matrix indices and bare parameters $Z,\lambda_{bare}$
identical as in our previous paper
\cite[eq.\ (5.5)]{Grosse:2016pob}:
\begin{align}
G_{|a^1|a^2|\dots |a^B|}
&= \frac{(-\lambda_{bare} Z^{\frac{3}{2}})}{ZH_{\under{a}^1\under{a}^1}}\Big\{
\frac{1}{V} \!\!\! \sum_{\under{n} \in \mathbb{N}^{D/2}_{\mathcal{N}}} \!\!
G_{|\under{a}^1\under{n}|\under{a}^2|\dots |\under{a}^B|}
+\sum_{\beta=2}^B G_{|\under{a}^1\under{a}^\beta \under{a}^\beta|
\under{a}^2| \stackrel{\beta}{\check{\dots\dots}}
  |\under{a}^B|}
+ \frac{1}{V^2} G_{|\under{a}^1|\under{a}^1|\under{a}^2|\dots |\under{a}^B|}
\nonumber
\\
&+ 2 G_{|\under{a}^1|}  G_{|\under{a}^1|\under{a}^2|\dots |\under{a}^B|}
+ \sum_{p=1}^{B-2} \sum_{2\leq i_1<\dots < i_p\leq B}
G_{|\under{a}^1|\under{a}^{i_1}|\dots |\under{a}^{i_p}|}
G_{|\under{a}^1|\under{a}^{j_1}|\dots |\under{a}^{j_{B-p-1}}|}\Big\},
\end{align}
where $\{j_1,\dots,j_{B-p-1}\}=\{2,\dots,B\}\setminus 
\{i_1,\dots ,i_p\}$ and $\stackrel{\beta}{\check{\dots\dots}}$ denotes 
the omission of $\under{a}^\beta$.
We multiply by $\frac{H_{\under{a}\under{a}}}{\lambda_r}
= \frac{2E_{\under{a}}}{\lambda_r}
=\frac{2F_{\under{a}}}{\lambda_r}+\nu$, see (\ref{EF}),
and bring
$-2 G_{|\under{a}^1|}  G_{|\under{a}^1|\under{a}^2|\dots |\under{a}^B|}$
to the lhs, thus reconstructing
the function $W_{|a^1|}$ defined in (\ref{2pointWard-b}) but
\emph{here shifted by $\nu\lambda_r$} compared with
\cite[eq.\ (5.4)]{Grosse:2016pob}.
Reducing the $(2{+}1{+}\dots{+}1)$-point function
by (\ref{GNB}), where the remarks of footnote \ref{fn2} apply, leads to
\begin{align}
(W_{|\under{a}^1|}+\nu\lambda_r)  &G_{|\under{a}^1|\under{a}^2|\dots |\under{a}^B|}
+\frac{\lambda_r^2}{V}
\sum_{\under{n}\in\mathbb{N}^{D/2}_{\mathcal{N}}}
\frac{G_{|\under{a}^1|\under{a}^2|\dots |\under{a}^B|}
-G_{|\under{n}|\under{a}^2|\dots |\under{a}^B|}}{(F_{\under{a}^1}^2-F_{\under{n}}^2)}
\label{G1B}
\\*
&=-\lambda_r
\sum_{\beta=2}^B G_{|\under{a}^1\under{a}^\beta \under{a}^\beta|\under{a}^2|
\stackrel{\beta}{\check{\dots\dots}}  |\under{a}^B|}
- \frac{\lambda_r}{V^2} G_{|\under{a}^1|\under{a}^1|\under{a}^2|\dots |\under{a}^B|}
\nonumber
\\*
&
- \lambda_r\sum_{p=1}^{B-2} \sum_{2\leq i_1<\dots < i_p\leq B}
G_{|\under{a}^1|\under{a}^{i_1}|\dots |\under{a}^{i_p}|}
G_{|\under{a}^1|\under{a}^{j_1}|\dots |\under{a}^{j_{B-p-1}}|}.
\nonumber
\end{align}

Taking the scaling limit $
\tilde{G}(x^1|\dots|x^B):=\mu^{(2-\frac{3}{2}B)D-B}
\lim_{\mathcal{N},V\to \infty}
G_{|\under{a}^1|\dots|\under{a}^B|}\big|_{|\under{a}^i| =
V^{\frac{2}{D}}\mu^2x^i}$ and transforming variables $x^i$ to $X^i$,
we essentially obtain \cite[eq.\ (5.7)]{Grosse:2016pob},
\begin{align}
&(W(X^1)+\tilde{\lambda}\tilde{\nu})
G(X^1|X\triangleleft^{\{2,\dots B\}})
+\frac{1}{2} \int_1^{\Xi} dT \rho(T)
\frac{G(X^1|X\triangleleft^{\{2,\dots B\}})-G(T|X\triangleleft^{\{2,\dots B\}})}{
(X-T)}
\nonumber
\\
&=-\tilde{\lambda}
\sum_{\beta=2}^B G\Big(X^1,X^\beta ,X^\beta|X\triangleleft^{\{2
\stackrel{\beta}{\check{\dots\dots}} B\}}\Big)
- \tilde{\lambda} \!\!\!
\sum_{{J\subset \{2,\dots,B\} \atop 1\leq |J| \leq B-2}} \!\!\!
G(X^1|X\triangleleft^J)
G(X^1|X\triangleleft^{\{2,\dots,B\}\setminus J}),
\label{G1B-lim}
\end{align}
where the measure $\rho(T)$ was defined in (\ref{Int_eq_field1})
and $G(X|Y\triangleleft^{\{i_1,\dots,i_p\}})
:=G(X|Y^{i_1}|\dots|Y^{i_p})$.
This equation was solved in our previous paper \cite{Grosse:2016pob}:
\begin{thm}
\label{Prop-1+1}
The scaling limit of the $(1{+}\dots{+}1)$-point function is given
by
\begin{align}
G(X|Y)&= \frac{4\tilde{\lambda}^2}{\sqrt{X+c}\cdot \sqrt{Y+c}\cdot
(\sqrt{X+c}+\sqrt{Y+c})^2},
\label{G1+1}
\end{align}
for $B=2$ and for $B\geq 3$ by
\begin{align}
G(X^1|\dots|X^B)&=
(-2\tilde{\lambda})^{3B-4}
\frac{d^{B-3}}{dt^{B-3}}
\Big(\frac{1}{(R(t))^{B-2} \sqrt{X^1{+}c{-}2t}^3
\cdots \sqrt{X^B{+}c {-}2t}^3}\Bigg)\Bigg|_{t=0},
\label{G1B-thm}
\\
R(t) &:= \lim_{\Xi\to \infty}\Big(\frac{1}{\sqrt{Z}}-
\int_1^\Xi \frac{dT
  \rho(T)}{\sqrt{T{+}c}} \frac{1}{(\sqrt{T{+}c}+\sqrt{T{+}c{-}2t})
\sqrt{T{+}c{-}2t}}\Big).
\end{align}
Here, $c(\tilde{\lambda},e)$ is defined in
\cite[eq.\ (4.15)]{Grosse:2016pob} for $D=2$, in
(\ref{MS-solution-a4}) for $D=4$ and in
(\ref{MS-solution-d}) for $D=6$. The wavefunctions renormalsiation
equals $Z=1$ for $D\in \{2,4\}$ and is given in (\ref{1/sqrtZ})
for $D=6$.
\end{thm}
Explicitly we have for $D=6$
\begin{align}
&R(t):=
\sqrt{1{+}c}-
\int_1^\infty \!\!\! dT\, \rho(T)\;
\frac{\begin{array}{r}\big\{
\sqrt{1{+}c}(2\sqrt{T{+}c}+\sqrt{1{+}c})
(\sqrt{T{+}c{-}2t}+\sqrt{T{+}c})
\qquad\qquad \\ +t(\sqrt{T{+}c{-}2t}+2\sqrt{T{+}c}) \big\}
\end{array}}{\sqrt{T{+}c}
(\sqrt{1{+}c}+\sqrt{T{+}c})^2(\sqrt{T{+}c}+\sqrt{T{+}c{-}2t})^2
\sqrt{T{+}c{-}2t}}.
\label{R6}
\end{align}

\section{Linearly spaced eigenvalues and 
Feynman graphs for $D=6$}

\label{sec:BPHZ}

\subsection{Expanding the exact result}

The noncommutative field theory model discussed in sec.\
\ref{sec:positivity} (see also \cite[sec.\ 6]{Grosse:2016pob})
translates to linearly spaced eigenvalues with $e(x)=x$ and $e'(x)=1$.
This yields $X=(2x+1)^2$ and
$\rho(T)=\frac{\tilde{\lambda}^2(\sqrt{T}-1)^2}{4\sqrt{T}}$.  The
integrals (\ref{MS-solution-c})+(\ref{MS-solution-d}) can be 
evaluated\footnote{We actually compute $\frac{1}{\sqrt{Z}}$ from
(\ref{1/sqrtZ}) and the integral (\ref{MS-solution-a}). A
primitive of the integrand is obtained by computer algebra; the limit 
$\Xi \to \infty$ is done by hand. 
Also identities of the type $\frac{\sqrt{1+c}+\sqrt{X+c}}{
c+\sqrt{X}+\sqrt{1+c}\sqrt{X+c}}
=\frac{1+\sqrt{X}}{\sqrt{X}\sqrt{1+c}+\sqrt{X+c}}$ are used.
The series expansions
(\ref{c6-pert}) and (\ref{G1-pert}) are found by computer algebra.}
for $\Xi\to \infty$:
\begin{prop} \label{prop_MS}
Let $D=6$. Equation  (\ref{Int_eq_field1}) with normalisation 
(\ref{normalisation-1}) is for eigenvalue functions $e(x)=x$ 
solved by:
\begin{align}
W(X)&=
\sqrt{X{+}c}\sqrt{1{+}c}-c +\frac{\tilde{\lambda}^2}{2}
\Big\{\sqrt{1{+}c}-\sqrt{1{+}X}+
 \log\Big(\frac{\sqrt{X+c} + \sqrt {1{+}c}}{
 2 (1 + \sqrt{1{+}c})}\Big)
\nonumber
\\*
&+ \frac{(1+X)}{2\sqrt{X}}
\log\Big( \frac{(\sqrt{X}+\sqrt{X{+}c})(1+\sqrt{X})
}{\sqrt{X}\sqrt{1{+}c}+\sqrt{X{+}c}}\Big)\Big\},
\label{WX-D=6}
\end{align}
where $c(\tilde{\lambda})$ is the inverse solution of 
\begin{align}
\tilde{\lambda}^2
&= \frac{(-4c)}{1
-2\sqrt{1+c}+2(1+c)\log(1+\frac{1}{\sqrt{1+c}})}.
\label{W1=1}
\end{align}
\end{prop}
The first terms of the inverse solution are
\begin{align}
c&= -\frac{2\log 2-1}{4} \tilde{\lambda}^2  
+\frac{ (2\log 2-1)(4\log 2 -3)}{32} \tilde{\lambda}^4  
\nonumber
\\
&- \frac{ (2\log 2-1)(35-94\log 2 +64(\log 2)^2)}{1024} \tilde{\lambda}^6  
+\mathcal{O}(\lambda^8).
\label{c6-pert}
\end{align}
This gives the following perturbative
expansion of the 1-point function:
\begin{align}
\tilde{G}(x)&:= \frac{W((2x+1)^2)-(2x+1)}{2\tilde{\lambda}}
\nonumber
\\
&= \frac{\tilde{\lambda}}{4(2x+1)} 
\big(2 (1 + x)^2 \log(1+x) - x(2 + 3 x)\big)
\nonumber
\\
& + \frac{\tilde{\lambda}^3}{16(2x+1)^3}
\big( x^3(2+3x)(2\log 2-1)^2 \big)
+ \mathcal{O}(\tilde{\lambda}^5).
\label{G1-pert}
\end{align}

\subsection{Perturbative expansion of the 
partition function}

On the other hand, expanding the original action as a formal power
series in $\lambda$ leads to a ribbon graph representation of 
$\log \mathcal{Z}[J]$. Ignoring the renormalisation constants, i.e.\
setting
$\mu_{bare}=\mu$, $\lambda_{bare}=\lambda$, 
$Z=1$, $\kappa=\nu=\zeta=0$ gives in the large-$(\mathcal{N},V)$ limit
`Feynman' rules for planar ribbon graphs with $B$
boundary components. We formulate these rules in an 
equivalent description \cite[sec.\ 2]{Grosse:2016pob}) by planar 
graphs $\Gamma$ on the 
2-sphere with two sorts of
vertices: any number of black (internal) vertices of valence 3, and
$B\geq 1$ white vertices $\{v_\beta\}_{\beta=1}^B$ (external vertices,
or punctures, or boundary components) of any valence $N_\beta\geq 1$.
Every face is required to have at most one white vertex (separation of
punctures). Faces with a white vertex are called external;
they are labelled by positive real numbers
$x^1_1,\dots,x^1_{N_1},\dots,x^B_1,\dots,x^B_{N_B}$ (the upper index
labels the unique white vertex of the face).  Faces without white
vertex are called internal; they are labelled by positive real numbers
$y_1,\dots, y_L$.  Such graphs are dual to triangulations of the
$B$-punctured sphere. To such a graph $\Gamma$ we assign an \emph{unrenormalised}
amplitude $\tilde{G}^\Lambda_{\Gamma}$ as follows:

\begin{itemize}\itemsep 0pt
\item Associate a weight $(-\tilde{\lambda})$ to each black vertex, 
weight $1$ to each 
white vertex of $\Gamma$.

\item Associate weight $\frac{1}{z_1+z_2+1}$ to an edge of $\Gamma$ 
separating
faces labelled by $z_1$ and $z_2$. These can be internal or external,
also $z_1=z_2$ can occur. 

\item Multiply these weights of $\Gamma$ 
and integrate over all internal face variables $y_1,\dots,y_L$ 
with measure $\frac{1}{2} y_i^2 \chi_{[0,\Lambda^2]}(y_i) \,dy_i$, where 
$\chi_{[0,\Lambda^2]}$ is the characteristic function of
$[0,\Lambda^2]$ and $d y_i$ the Lebesgue measure on $\mathbb{R}_+$.

\item The result is a function 
$\tilde{G}^\Lambda_\Gamma(x^1_1,\dots,x^1_{N_1}|\dots|x^B_1,\dots,x^B_{N_B})$ of 
the external face variables. 

\end{itemize}

We list the simplest graphs contributing to the 1-point function 
$\tilde{G}^{\Lambda}(x)$  and
their unrenormalised integrals:
\begin{align}
\Gamma_1&=\parbox{20mm}{\begin{picture}(20,10)
\put(15,5){\circle{10}}
\put(0.5,5){\line(1,0){9.5}}
\put(-1,4){\mbox{$\circ$}}
\put(9,4){\textbullet}
\put(3,6.5){\mbox{\small$x$}}
\put(13,5){\mbox{\small$y_1$}}
\end{picture}} 
&
\tilde{G}_{\Gamma_1}^{\Lambda}(x)&= 
\frac{(-\tilde{\lambda})}{2x+1} 
\int_0^{\Lambda^2} 
\frac{dy_1\;y_1^2/2}{x+y_1+1},
\label{gaphs-G1}
\\
\Gamma_2&=\parbox{22mm}{~\begin{picture}(22,8)
\put(0.5,4){\line(1,0){4.5}}
\put(8.5,4){\circle{7}}
\put(11.5,4){\line(1,0){3}}
\put(18.5,4){\circle{7}}
\put(7,4){\mbox{\small$y_1$}}
\put(17,4){\mbox{\small$y_2$}}
\put(0,6){\mbox{\small$x$}}
\put(4,3){\textbullet}
\put(11,3){\textbullet}
\put(14,3){\textbullet}
\put(-1,3){\mbox{$\circ$}}
\end{picture}}
&
\tilde{G}_{\Gamma_2}^{\Lambda}(x)&= 
 \frac{(-\tilde{\lambda})^3}{(2x+1)^2}
\int_0^{\Lambda^2} 
\frac{dy_1\;y_1^2/2}{(x{+}y_1{+}1)^2}
\int_0^{\Lambda^2} 
\frac{dy_2\;y_2^2/2}{x{+}y_2{+}1},
\nonumber
\\
\Gamma_3&=
\parbox{25mm}{\begin{picture}(25,16)
\put(0.5,8){\line(1,0){5.5}}
\put(15,8){\oval(18,16)}
\put(16,8){\circle{8}}
\put(24,8){\line(-1,0){4}}
\put(14,8){\mbox{\small$y_2$}}
\put(9,12){\mbox{\small$y_1$}}
\put(1,10){\mbox{\small$x$}}
\put(5,7){\textbullet}
\put(23,7){\textbullet}
\put(19.4,7){\textbullet}
\put(-1,7){\mbox{$\circ$}}
\end{picture}}
&
\tilde{G}_{\Gamma_3}^{\Lambda}(x)&= 
 \frac{(-\tilde{\lambda})^3}{2x+1}
\int_0^{\Lambda^2} \frac{dy_1\;y_1^2/2}{(x{+}y_1{+}1)^2(2y_1+1)}
\int_0^{\Lambda^2} \frac{dy_2\;y_2^2/2}{y_1{+}y_2{+}1},
\nonumber
\\
\Gamma_4&=\parbox{20mm}{\begin{picture}(20,16)
\put(0.5,8){\line(1,0){7.5}}
\put(8,8){\line(2,1){4}}
\put(8,8){\line(2,-1){4}}
\put(14.5,12.5){\circle{7}}
\put(14.5,3.5){\circle{7}}
\put(13,3){\mbox{\small$y_1$}}
\put(13,12){\mbox{\small$y_2$}}
\put(1,10){\mbox{\small$x$}}
\put(7,7){\textbullet}
\put(11.2,9){\textbullet}
\put(11.2,5){\textbullet}
\put(-1,7){\mbox{$\circ$}}
\end{picture}}
&
\tilde{G}_{\Gamma_4}^{\Lambda}(x)&= 
\frac{(-\tilde{\lambda})^3}{(2x+1)^3}
\int_0^{\Lambda^2} \frac{dy_1\;y_1^2/2}{x{+}y_1{+}1}
\int_0^{\Lambda^2} \frac{dy_2\;y_2^2/2}{x{+}y_2{+}1},
\nonumber
\\[1ex]
\Gamma_5&=
\parbox{25mm}{\begin{picture}(25,16)
\put(0.5,8){\line(1,0){5.5}}
\put(15,8){\oval(18,16)}
\put(15,0){\line(0,1){16}}
\put(18,8){\mbox{\small$y_2$}}
\put(9,8){\mbox{\small$y_1$}}
\put(1,10){\mbox{\small$x$}}
\put(5,7){\textbullet}
\put(14,-1){\textbullet}
\put(14,15){\textbullet}
\put(-1,7){\mbox{$\circ$}}
\end{picture}}
&
\tilde{G}_{\Gamma_5}^{\Lambda}(x)&= 
\frac{(-\tilde{\lambda})^3}{(2x+1)}
\int_0^{\Lambda^2} \int_0^{\Lambda^2} 
\frac{dy_1dy_2 \;(y_1^2/2)(y_2^2/2)
}{(x{+}y_1{+}1)^2(y_1{+}y_2{+}1)
(x{+}y_2{+}1)}.
\nonumber
\end{align}
Clearly, all these integrals diverge for $\Lambda\to
\infty$. By employing the standard techniques of perturbative
renormalisation theory it should be possible to prove that there exist
formal power series 
$\mu_{bare}=\mu+\mathcal{O}(\lambda)$, 
$\lambda_{bare}=\lambda+\mathcal{O}(\lambda)$, 
$Z=1+\mathcal{O}(\lambda)$, $\kappa=\mathcal{O}(\lambda)$,
$\nu=\mathcal{O}(\lambda)$,
$\zeta=\mathcal{O}(\lambda)$, divergent at $\Lambda\to \infty$, 
such that all matrix correlation functions are finite to all orders in
perturbation theory. Moreover, the value of these correlation
functions is expected to be uniquely determined by normalisation
conditions e.g.\ on
\begin{align}
\tilde{G}(0),\quad
 (\partial \tilde{G})(0), \quad(\partial^2 \tilde{G})(0), \quad
\tilde{G}(0,0),\quad (\partial \tilde{G})(0,0), \quad\tilde{G}(0,0,0).
\end{align}

\subsection{Zimmermann's forest formula for 
ribbon graphs}

A key step in proving these claims is the \emph{forest formula} of
Wolfhart Zimmermann \cite{Zimmermann:1969jj}. Streamlining previous
(and very essential) work of Klaus Hepp \cite{Hepp:1966eg} on the
Bogoliubov-Parasiuk renormalisation description
\cite{Bogoliubov:1957??}, Zimmermann proved that those formal power
series $\mu_{bare}(\lambda)$, $\lambda_{bare}(\lambda)$, $Z(\lambda)$,
$\kappa(\lambda)$, $\nu(\lambda)$, $\zeta(\lambda)$ which enforce the
normalisation conditions (\ref{normalisation}) simply amount to a
well-defined modification, for every ribbon graph $\Gamma$, of the
\emph{integrand} $I_\Gamma $ of $\Gamma$ (product of vertex weights
and edge weights). The modified integrand $\mathcal{R}(I_\Gamma)$ is
Lebesgue-integrable over the whole space and by Fubini's theorem can
be unambiguously integrated in any order.

We describe this modification $\mathcal{R}$ for the $\Phi^3_6$-matrix
model under consideration. The situation is far easier than in the
papers by Hepp and Zimmermann because in Euclidean space there is no
need to discuss the $\mathrm{i}\epsilon$-limit to tempered
distributions. Moreover, the globally assigned face variables give
unique edge weights, in contrast to choices of momentum routings in
ordinary Feynman graphs that make recursive substitution operations
necessary.

Let $\mathcal{B}_\Gamma,\mathcal{F}_\Gamma$ be the set of external and
internal faces of $\Gamma$, respectively.  A \emph{ribbon subgraph}
$\gamma \subset \Gamma$ consists of a subset
$\mathcal{F}_\gamma\subset \mathcal{F}_\Gamma$ together with all edges
and vertices bordering $\mathcal{F}_\gamma$, and not more, such that
the (thus defined) set of edges and vertices of $\gamma$ is connected.
The ribbon subgraph defines a unique subset $\mathcal{E}_\gamma
\subset \mathcal{B}_\Gamma\cup \mathcal{F}_\Gamma \setminus
\mathcal{F}_\gamma$ of neighbouring faces to $\gamma$, i.e.\ any
element of $\mathcal{E}_\gamma$ has a common edge with $\gamma$.  We
let $B_\gamma$ be the number of connected components of
$\mathcal{E}_\gamma$. This number is most conveniently identified when
drawing $\Gamma$ on the 2-sphere where $\mathcal{E}_\gamma$ will
partition into $B_\gamma$ disjoint regions.

We let $f(\gamma)$ be the set of those 
face variables $\{x,y\}$ which label the faces in $\mathcal{E}_\gamma$.
Then for a rational function $r_\gamma(f(\gamma),y_1,\dots,y_l)$, where 
$y_1,\dots,y_l$ label the faces of $\gamma$, we let 
$(T^\omega_{f(\gamma)} r_\gamma)(f(\gamma),y_1,\dots,y_l)$ be 
the order-$\omega$ Taylor polynomial of $r_\gamma$ with respect 
to the variables $f(\gamma)$. We let 
$T^\omega_{f(\gamma)} r_\gamma \equiv 0$ for $\omega <0$.

A forest $\mathcal{U}_\Gamma$ in $\Gamma$ is a collection of ribbon
subgraphs $\gamma_1,\dots,\gamma_l$ such that for any pair
$\gamma_i,\gamma_j$
\[
\text{either $\gamma_i \subset \gamma_j$, \quad or 
$\gamma_j \subset \gamma_i$, \quad  or 
$\gamma_i \cap \gamma_j=\varnothing$.}
\]
Here $\gamma_i \cap \gamma_j=\varnothing$ means complete disjointness
in the sense that $\gamma_i, \gamma_j$ do not have any common edges
or vertices. Similarly, $\gamma_j \subset \gamma_i$ means that 
all faces, edges and vertices of $\gamma_j$ also belong to $\gamma_i$.
We admit $\mathcal{U}_\Gamma =\varnothing$, but note that $\Gamma$ itself cannot belong to $\mathcal{U}_\Gamma$ because $\Gamma$ has purely 
external edges which are not part of a ribbon subgraph. 

Consequently, a forest is endowed with a partial 
order which we symbolise by disjoint trees (hence the name). 
The tree structure identifies for any $\gamma_i\in \mathcal{U}_\Gamma$ a 
unique set $o(\gamma_i)= \gamma_{i_1}\cup \dots \cup 
\gamma_{i_k}$ of offsprings, 
i.e.\ mutually disjoint ribbon subgraphs $\gamma_{i_h} \subset \gamma_i$ 
such that for any other $\gamma_j \in \mathcal{U}_{\Gamma} \setminus 
(\gamma_i \cup o(\gamma_i))$ either $\gamma_j \subset o(\gamma_i)$,
or $\gamma_i \subset \gamma_j$, or
$\gamma_i \cap \gamma_j=\varnothing$.
A ribbon subgraph $\gamma$ with $o(\gamma)=\varnothing$ is called a leaf.

Let $I_\Gamma$ be the integrand encoded in a ribbon graph, given as
the product of weights of edges of $\Gamma$ (the constant vertex
weights play no r\^ole). 
A forest 
$\mathcal{U}_\Gamma$ defines a partition of the integrand $I_\Gamma$ into 
\begin{align}
I_\Gamma= 
I_{\Gamma\setminus \mathcal{U}_\Gamma} 
\prod_{\gamma \in \mathcal{U}_\Gamma} I_{\gamma\setminus o(\gamma)},\qquad 
\prod_{\gamma \in \varnothing} I_\varnothing =1.
\end{align}

Then the following holds:
\begin{thm}[after Zimmermann]
The formal power series
$\mu_{bare}(\lambda)$, 
$\lambda_{bare}(\lambda)$, 
$Z(\lambda)$, $\kappa(\lambda)$,
$\nu(\lambda)$,
$\zeta(\lambda)$ which enforce the 
normalisation conditions (\ref{normalisation}) amount for any 
ribbon graph $\Gamma$ to replace the integrand $I_\Gamma$ as follows:
\begin{align}
I_\Gamma 
&\mapsto \mathcal{R}(I_\Gamma):= \sum_{\mathcal{U}_\Gamma} 
I_{\Gamma\setminus \mathcal{U}_\Gamma} 
\prod_{\gamma \in \mathcal{U}_\Gamma} 
\big(-T^{\omega(\gamma)}_{f(\gamma)} I_{\gamma\setminus o(\gamma)}\big),
\end{align}
where the sum is over all forests $\mathcal{U}_\Gamma$ of $\Gamma$ including 
the empty forest $\varnothing$. For the planar sector of the 
$\Phi^3_D$-matrix model, the degree is
defined as $\omega(\gamma):=\frac{D}{2}(2-B_\gamma)-N_\gamma$, 
where $N_\gamma$ is
the number of edges of $\Gamma$ which connect (within 
$\Gamma$) to
vertices of $\gamma$ but are not in $\gamma$ itself.
\end{thm}

\subsection{Forest formula applied to $1$-point function}

We exemplify Zimmermann's rules for the ribbon graphs given in
(\ref{gaphs-G1}). The graph $\Gamma_1$ has two forests
$\mathcal{U}_{\Gamma_1}=\varnothing$ and
$\mathcal{U}_{\Gamma_1}=\{\gamma_1\}$, where $\gamma_1$ is the ribbon
subgraph consisting of the face $y_1$ and its bordering line and
vertex. Taking $N_{\gamma_1}=1$ and $B_{\gamma_1}=1$ into account, we
have to replace
\begin{align}
\frac{1}{x+y_1+1} \mapsto 
\underbrace{\frac{1}{x+y_1+1}}_{\text{for }\mathcal{U}_{\Gamma_1}=\varnothing}
+\underbrace{(-T^2_x)\Big(\frac{1}{x+y_1+1}\Big)}_{\text{for }
\mathcal{U}_{\Gamma_1}=\{\gamma_1\}}.
\end{align}
With 
$(-T^2_x)(\frac{1}{x+y_1+1})
= -\frac{1}{y_1+1}+\frac{x}{(1+y_1)^2}-\frac{x^2}{(1+y_1)^3}$ we 
compute the renormalised amplitude of the ribbon graph $\Gamma_1$ to
\begin{align}
\tilde{G}_{\Gamma_1}(x)
&= \frac{(-\tilde{\lambda})}{2x+1} 
\int_0^\infty \frac{y_1^2 dy_1}{2} \Big(
\frac{1}{x+y_1+1}-
\frac{1}{y_1+1}+
\frac{x}{(y_1+1)^2}-
\frac{x^2}{(y_1+1)^3}\Big)
\nonumber
\\
&=\frac{(-\tilde{\lambda})}{4(2x+1)} 
\big(x (2 + 3 x) - 2 (1 + x)^2 \log(1+x)\big).
\label{G1-pert-1}
\end{align}
We confirm the perfect agreement with the $\lambda$-expansion 
(\ref{G1-pert}) of the exact formula!

The next graph $\Gamma_2$ has four forests 
$U_{\Gamma_2}=\varnothing$, 
$U_{\Gamma_2}=\{\gamma_1\}$, 
$U_{\Gamma_2}=\{\gamma_2\}$, 
$U_{\Gamma_2}=\{\gamma_1,\gamma_2\}$.
By $\gamma_i$ we denote the ribbon subgraph with face labelled $y_i$ 
together with its bordering edges and vertices. 
Note that the faces labelled $y_1,y_2$ together do not give rise 
to a ribbon subgraph because its edge+vertex set would be disconnected.
We have $N_{\gamma_1}=2$,
$N_{\gamma_2}=1$ and $B_{\gamma_1}=B_{\gamma_2}=1$. We have
$f(\gamma_1)=f(\gamma_2)=\{x\}$.
Hence, the Taylor subtraction operator for $\gamma_1$ is 
$(-T^1_{x})(\frac{1}{(x+y_1+1)^2})=-\frac{1}{(y_1+1)^2}+
\frac{2x}{(y_1+1)^3}$,
and $(-T^2_{x})(\frac{1}{(x+y_2+1)})$ is analogous to $\Gamma_1$. 
Then Zimmermann's forest formula factors as follows:
\begin{align}
\tilde{G}_{\Gamma_2}(x)
&:= \frac{(-\tilde{\lambda})^3}{(2x+1)^2}
\int_0^\infty \frac{y_1^2 dy_1}{2}
\Big(\frac{1}{(x{+}y_1{+}1)^2}
-\frac{1}{(y_1{+}1)^2}
+\frac{2x}{(y_1{+}1)^3}
\Big)
\nonumber
\\
& \hspace*{25mm} \times 
\int_0^\infty \frac{y_2^2 dy_2}{2}
\Big(
\frac{1}{x+y_2+1}-
\frac{1}{y_2+1}+
\frac{x}{(y_2+1)^2}-
\frac{x^2}{(y_2+1)^3}\Big)
\nonumber
\\
&= \frac{\tilde{\lambda}^3}{4(2x+1)^2}
\Big((1+x)\log(1+x) -x\Big)
\Big( 2 (1 + x)^2 \log(1+x)-x (2 + 3 x) \Big).
\end{align}

In $\Gamma_3$ we let the ribbon subgraph $\gamma_2$ as before and
define $\gamma_{12}$ as consisting of \emph{both} faces labelled by
$y_1$ and $y_2$ together with all bordering edges and vertices. There
is also the ribbon subgraph $\gamma_1$ consisting of $y_1$ but not
$\gamma_2$. Then $\mathcal{E}_{\gamma_2}$ has $B_{\gamma_2}=2$
components, namely the disconnected faces labelled by $y_2$ and by
$x$. In principle we could include $\gamma_1$ in the forest formula,
but $\omega(\gamma_1)=-1$ forces the Taylor operator $T$ to vanish.
Therefore, we can work with the reduced set of forests
$U_\Gamma=\varnothing$, $U_\Gamma= \{\gamma_2\}$, $U_\Gamma=
\{\gamma_{12}\}$ and $U_\Gamma= \{\gamma_{12},\gamma_2\}$. The
external face variables are $f(\gamma_2)=\{y_1\}$ and 
$f(\gamma_{12})=\{x\}$.  The
resulting analytic contribution is\footnote{We compute the primitive
  by computer algebra and take the limit $\Lambda^2\to \infty$ by
  hand, where polylogarithmic identities are employed.}
\begin{align}
\tilde{G}_{\Gamma_3}(x)
&= \frac{(-\tilde{\lambda})^3}{2x+1}
\int_0^\infty \frac{y_1^2 dy_1}{2}
\Big(\frac{1}{(x{+}y_1{+}1)^2}
-\frac{1}{(y_1{+}1)^2}
+\frac{2x}{(y_1{+}1)^3}
-\frac{3 x^2}{(y_1{+}1)^4}
\Big)
\nonumber
\\*
& \hspace*{25mm} \times \frac{1}{2y_1+1} 
\int_0^\infty \frac{y_2^2 dy_2}{2}
\Big(
\frac{1}{y_1{+}y_2{+}1}-
\frac{1}{y_2{+}1}+
\frac{y_1}{(y_2{+}1)^2}-
\frac{(y_1)^2}{(y_2{+}1)^3}\Big)
\nonumber
\\
&=  \frac{\tilde{\lambda}^3}{4(2x{+}1)^3}\Big\{ 
x^3 (2 {+} 3 x) \Big( \frac{(1 - 2 \log  2)^2}{4}-\frac{\pi^2}{2} \Big)  
+(1 {+} x)^2 (2 {+} 7 x {+} 7 x^2) \log(1 {+} x)
\nonumber
\\
& 
+x (1 {+} x) (1 {+} 3 x {+} 3 x^2) \Big((\log(1{+} x))^2 
- 2 \log(1 {+} x)\log x +
2 \mathrm{Li}_2\Big(\frac{1}{1{+}x}\Big)\Big)
+\frac{x^4}{4}
\Big\}
\nonumber
\\
&
-\frac{\tilde{\lambda}^3}{2(2x+1)}\Big(\frac{\pi^2}{6}+1\Big)x.
\end{align}
We see that the whole number-theoretic features of quantum field
theory are reproduced!

The graph $\Gamma_4$ poses no difficulty:
\begin{align}
\tilde{G}_{\Gamma_4}(x)
&= \frac{(-\tilde{\lambda})^3}{(2x+1)^3}
\int_0^\infty \frac{y_1^2 dy_1}{2}
\Big(\frac{1}{x{+}y_1{+}1}
-\frac{1}{(y_1{+}1)}
+\frac{x}{(y_1{+}1)^2}
-\frac{x^2}{(y_1{+}1)^3}
\Big)
\nonumber
\\
& \hspace*{25mm} \times 
\int_0^\infty \frac{y_2^2 dy_2}{2}
\Big(
\frac{1}{x+y_2+1}-
\frac{1}{y_2+1}+
\frac{x}{(y_2+1)^2}-
\frac{x^2}{(y_2+1)^3}\Big)
\nonumber
\\
&= \frac{-\tilde{\lambda}^3}{16(2x+1)^3} 
\Big( 2 (1 + x)^2 \log(1+x)-x (2 + 3 x) \Big)^2.
\end{align}

The graph $\Gamma_5$ shows a new quality: the overlapping divergence 
made of the ribbon subgraphs $\gamma_1$ and $\gamma_2$ which share 
a common edge. Overlapping divergences were a problem in the first 
days of quantum field theory. Now with the forest formula at 
disposal there is nothing to worry. The definition simply forbids  
$\gamma_1$ and $\gamma_2$ in the same forest. The following forests remain:
\[
\mathcal{U}_{\Gamma_5}=\varnothing,\quad 
\mathcal{U}_{\Gamma_5}=\{\gamma_1\},\quad 
\mathcal{U}_{\Gamma_5}=\{\gamma_2\},\quad 
\mathcal{U}_{\Gamma_5}=\{\gamma_{12}\},\quad 
\mathcal{U}_{\Gamma_5}=\{\gamma_{12},\gamma_1\},\quad 
\mathcal{U}_{\Gamma_5}=\{\gamma_{12},\gamma_2\}.
\]
We have $N_{\gamma_{12}}=1$, $N_{\gamma_{1}}=3$,
$N_{\gamma_{2}}=2$ and   
$B_{\gamma_{12}}=B_{\gamma_{1}}=B_{\gamma_{2}}=1$, giving 
$\omega(\gamma_{12})=2$, $\omega(\gamma_{1})=0$,
$\omega(\gamma_{2})=1$.
The face variables are $f(\gamma_{12})=\{x\}$,
$f(\gamma_{1})=\{x,y_2\}$,
$f(\gamma_{2})=\{x,y_1\}$. It follows:
\begin{align}
\tilde{G}_{\Gamma_5}(x)
&=
\frac{(-\tilde{\lambda})^3}{(2x+1)}
\int_0^\infty \frac{y_1^2 dy_1}{2}
\int_0^\infty \frac{y_2^2 dy_2}{2}\Bigg\{
\underbrace{\frac{1}{(x{+}y_1{+}1)^2}
\frac{1}{(y_1{+}y_2{+}1)}
\frac{1}{(x{+}y_2{+}1)}}_{I_{\gamma_{12}}}
\nonumber
\\
& \hspace*{20mm}
+
\underbrace{\frac{1}{x{+}y_2{+}1}}_{I_{\gamma_{12}\setminus \gamma_1}}
\underbrace{\Big(- \frac{1}{(y_1{+}1)^3}\Big)}_{(-T^0_{x,y_2})( I_{\gamma_1})}
+
\underbrace{\frac{1}{(x{+}y_1{+}1)^2}}_{I_{\gamma_{12}\setminus \gamma_2}}
\underbrace{\Big(-\frac{1}{(y_2{+}1)^2}
+\frac{y_1+x}{(y_2{+}1)^3}
\Big)}_{(-T^1_{x,y_1})(I_{\gamma_2})}
\nonumber
\\
&  \hspace*{20mm}
\underbrace{\begin{array}{@{}r}\displaystyle 
+ \frac{1}{(y_1{+}y_2{+}1)}\Big(-\frac{1}{(y_1{+}1)^2
(y_2{+}1)}
+ \frac{2x}{(y_1{+}1)^3(y_2{+}1)}
+\frac{x}{(y_1{+}1)^2(y_2{+}1)^2}
\\
\displaystyle -  \frac{3x^2}{(y_1{+}1)^4(y_2{+}1)}
-\frac{x^2}{(y_1{+}1)^2(y_2{+}1)^3}
-\frac{2x^2}{(y_1{+}1)^3(y_2{+}1)^2}
\Big)
\end{array}}_{(-T^2_{x})(I_{\gamma_{12}})}
\nonumber
\\
&  \hspace*{20mm}
+\underbrace{\Big(- \frac{1}{(y_1{+}1)^3}\Big)
\Big(-\frac{1}{y_2{+}1}
+\frac{x}{(y_2{+}1)^2}
-\frac{x^2}{(y_2{+}1)^3}
\Big)
}_{
(-T^2_{x})\text{\normalsize(}
I_{\gamma_{12}\setminus \gamma_1}
(-T^0_{x,y_2})(I_{\gamma_1})
\text{\normalsize)}}
\nonumber
\\
&  \hspace*{20mm}
\underbrace{\begin{array}{@{}r}\displaystyle 
+\Big(\Big(-\frac{1}{(y_1{+}1)^2}
+\frac{2x}{(y_1{+}1)^3}
-\frac{3x^2}{(y_1{+}1)^4}
\Big)
\Big(-\frac{1}{(y_2{+}1)^2}
+\frac{y_1}{(y_2{+}1)^3}
\Big)
\\\displaystyle 
+ 
\Big(-\frac{1}{(y_1{+}1)^2}
+\frac{2x}{(y_1{+}1)^3}
\Big)
\Big(
\frac{x}{(y_2{+}1)^3}
\Big)
\Big)
\end{array}
}_{
(-T^2_{x})\text{\normalsize(}
I_{\gamma_{12}\setminus \gamma_2}
(-T^1_{x,y_1})(I_{\gamma_2})
\text{\normalsize)}}
\Bigg\}
\nonumber
\\
&=  \frac{-\tilde{\lambda}^3}{4(2x{+}1)^3}\Big\{ 
(x {+} 1) (2 x {+} 1) (3 x {+} 2) \log(1 {+} x) + (x {+} 1)^3 (3 x {+} 1) 
(\log( 1 {+} x))^2
\nonumber
\\
& 
+x (1 {+} x) (1 {+} 3 x {+} 3 x^2) 
\Big((\log(1{+} x))^2 - 2 \log(1 {+} x)\log x +
2 \mathrm{Li}_2\Big(\frac{1}{1{+}x}\Big)\Big)
\nonumber
\\
&
-x^3 (2 {+} 3 x) \frac{\pi^2}{2} 
\Big\}
+\frac{\tilde{\lambda}^3}{2(2x{+}1)}\Big(\frac{\pi^2}{6}+1
-\frac{x}{2}\Big)x.
\end{align}

The sum is indeed the $\lambda^3$-part of (\ref{G1-pert}):
\begin{align}
\tilde{G}_{\Gamma_2}(x)+
\tilde{G}_{\Gamma_3}(x)+
\tilde{G}_{\Gamma_4}(x)+
\tilde{G}_{\Gamma_5}(x)=
\frac{\tilde{\lambda}^3}{16(2x+1)^3}
x^3 (2 + 3 x) (1 - 2 \log  2)^2.
\end{align}
This coincidence looks remarkable, but it isn't. It is the 
unavoidable consequence that two correct theorems about the 
same object must agree.

\subsection{Renormalons}

Renormalised amplitudes typically involve a logarithmic dependence on
the external parameters. Our model is no exception, see
(\ref{G1-pert-1}). In just renormalisable QFT models one finds that
$n$-fold insertion of these subgraphs into a superficially convergent
graph produces an unprotected amplification of the logarithms which
let the amplitude of the renormalised graph grow with
$\mathcal{O}(n!)$. Since there are also $\mathcal{O}(n!)$ graphs
contributing to order $n$, there is no hope of a Borel-summable
perturbation series. This phenomenon is called the \emph{renormalon
  problem}. It is an artefact of the strictly local Taylor subtraction
at $0$. Constructive renormalisation theory \cite{Rivasseau:1991ub} 
avoids the problem by a
reorganisation into an effective series of infinitely many (but
related) coupling constants. On the downside we loose the nice 
forest formula so that it is hard to compute the amplitudes in practice.

In this subsection we convince ourselves that the $\Phi^3_6$-matrix
model under consideration has, in its perturbative expansion, graphs
showing the renormalon problem. Consider the finite $(N=4,\
B=1)$-graph with a single internal face labelled $y$. We replace the 
edge between 
two black vertices by a chain of $n$ vertices and $n+1$ edges and attach to
every additional black vertex and towards the inner face 
the simplest 1-point function $\Gamma_1$: 
\begin{align}
\Gamma_r=~~&\parbox{50mm}{\begin{picture}(50,25)
\put(26,17){\oval(40,16)}
\qbezier(6,17)(0,17)(0,11)
\qbezier(0,11)(0,0)(25.3,0)
\qbezier(46,17)(52,17)(52,11)
\qbezier(52,11)(52,0)(26.6,0)
\put(14,25){\line(0,-1){5}}
\put(23,25){\line(0,-1){5}}
\put(38,25){\line(0,-1){5}}
\put(25.5,0.5){\line(-1,1){8.5}}
\put(26.5,0.5){\line(1,1){8.5}}
\put(5,16){\textbullet}
\put(45,16){\textbullet}
\put(13,24){\textbullet}
\put(22,24){\textbullet}
\put(37,24){\textbullet}
\put(13,19){\textbullet}
\put(22,19){\textbullet}
\put(37,19){\textbullet}
\put(16,8){\textbullet}
\put(34,8){\textbullet}
\put(14,16.5){\circle{7}}
\put(23,16.5){\circle{7}}
\put(38,16.5){\circle{7}}
\put(26,22){\dots\dots}
\put(25,-1){\mbox{$\circ$}}
\put(2.5,8){\mbox{\small$x_2$}}
\put(1.5,21){\mbox{\small$x_1$}}
\put(45,8){\mbox{\small$x_4$}}
\put(24,4){\mbox{\small$x_3$}}
\put(29,11){\mbox{\small$y$}}
\put(12.4,16.5){\mbox{\small$y_1$}}
\put(21.4,16.5){\mbox{\small$y_2$}}
\put(36.4,16.5){\mbox{\small$y_n$}}
\end{picture}}
\end{align}
Taking the renormalised 
amplitude (\ref{G1-pert-1}) into account, the amplitude of the 
total ribbon graph becomes in the simplified case that all external 
face variables are equal,
\begin{align}
\tilde{G}_{\Gamma_r}(x,x,x,x)
&=\frac{(-\tilde{\lambda})^{4+n} \cdot \tilde{\lambda}^n}{
(2x{+}1)^4}
\int_0^\infty \frac{y^2 dy}{2} 
\frac{1}{(x{+}y{+}1)^{n+4}}
\Big(\frac{2(1{+}y)^2\log(1{+}y)-y(2{+}3y)}{4(2y{+}1)}\Big)^n.
\end{align}
For large $y$ the integral behaves as
\[
\tilde{G}_{\Gamma_r}(x,x,x,x)
\sim
\frac{(-1)^n \tilde{\lambda}^{4+2n}}{
2\cdot 4^n (2x{+}1)^4}
\int_R^\infty \frac{dy}{y^2} (\log y)^n 
=
\frac{(-1)^n \tilde{\lambda}^{4+2n}}{
2\cdot 4^n (2x{+}1)^4}
\underbrace{\int_{\log R}^\infty dt \; e^{-t} t^n}_{\sim  n!}.
\]
This is the renormalon problem for a single graph.

On the other hand we know that the exact formula is analytic in
$\tilde{\lambda}$. There is a subtle cancellation
between different graphs, similar to the $\log(1+x)$ contributions of
$\tilde{G}_{\Gamma_{2\dots 5} }$ which all cancelled in the sum. In a
certain sense, this cancellation is another instance of the
integrability of the model.

\section{From $\phi^{\star 3}_D$ model on 
Moyal space to 
Schwinger functions on $\mathbb{R}^D$}

\label{sec:positivity}

This section parallels the treatment of $\phi^{\star 4}_4$ in
\cite{Grosse:2012uv, Grosse:2013iva} and $\phi^{\star 3}_2$ in 
\cite{Grosse:2016pob} where more details are given.
The (unrenormalised) $\phi^{\star 3}_D$-model on Moyal-deformed 
Euclidean space with harmonic propagation is defined by the action
\begin{align}
S[\phi]:= \int_{\mathbb{R}^D} \frac{d\xi}{(8\pi)^{D/2}}
\Big(\frac{1}{2}
\phi \star (-\Delta + \|4 \Theta^{-1} \cdot \xi\|^2 +\mu^2) \phi
+\frac{\lambda}{3} \phi\star\phi\star\phi\Big)(\xi),
\label{action-Moyal}
\end{align}
where $\star$ denotes the associative, noncommutative Moyal product
parametrised by a skew-symmetric matrix $\Theta$.
The Moyal space possesses a convenient matrix basis 
$f_{\under{m}\under{n}}(x)$ labelled by pairs of $\frac{D}{2}$-tupels 
$\under{m}=(m_1,\dots, m_{D/2})$ for which we set  
$|\under{m}|:=m_1+\dots+m_{D/2}$. The matrix basis
satisfies
$(f_{\under{k}\under{l}}\star f_{\under{m}\under{n}})(\xi)=\delta_{\under{m}\under{l}}
f_{\under{k}\under{n}}(\xi)$ and $\int_{\mathbb{R}^D}d\xi
\;f_{\under{m}\under{n}}(\xi)=\sqrt{|\det (2\pi\Theta)|}
\delta_{\under{m}\under{n}}$. A convenient
regularisation consists in
restricting the fields $\phi$ to those with finite expansion
$\phi(\xi)=\sum_{\under{m},\under{n} \in
  \mathbb{N}_{\mathcal{N}}^{D/2}} \Phi_{\under{m}\under{n}}
f_{\under{m}\under{n}}(\xi)$, where we recall 
$\mathbb{N}_{\mathcal{N}}^{D/2}:=\{\under{m}\in \mathbb{N}^{D/2}\;:~
|\under{m}|\leq  \mathcal{N}\}$.
Then (\ref{action-Moyal}) takes
precisely the form of our starting point (\ref{Hmn}) 
with undone renormalisation ($\kappa=\nu=\zeta=0$, $Z=1$,
$\lambda_{bare}=\lambda$, $\mu_{bare}=\mu$) and  
identification 
\begin{align}
V=\frac{1}{2^D} \sqrt{|\det \Theta|},\qquad 
E_{\under{n}}= \frac{|\under{n}|}{V^{2/D}}+\frac{\mu^2}{2}
=\mu^2\Big(\frac{1}{2}+\frac{\under{|n|}}{\mu^2 V^{2/D}}\Big).
\end{align}
Comparing with (\ref{En}), the Moyal space leads to a linear 
eigenvalue function $e(x)=x$.

According to \cite{Grosse:2013iva, Grosse:2016pob} the
large-($\mathcal{N},V$) limit of the matrix model induces together
with the convention $\frac{\delta J_{\under{m}\under{n}}}{\delta
  J(\xi)}:= \mu^D f_{\under{m}\under{n}}(\xi)$ the following relation
for connected Schwinger functions on extreme Moyal space
(where $s_\beta:=\sum_{i=1}^{\beta-1} N_i$):
\begin{align}
S_c(\mu \xi_1,{\dots},\mu \xi_N)
&= \!\!\!\!\!
\sum_{\stackrel{N_1+\dots+N_B=N}{N_\beta \text{ even}}}\sum_{\sigma  \in \mathcal{S}_N}
\prod_{\beta=1}^B \bigg(\frac{2^{\frac{D N_\beta}{2}}}{N_\beta}
\int_{\mathbb{R}^D} \! \frac{dp^\beta}{(2\pi\mu^2)^{\frac{D}{2}}} e^{\mathrm{i}\langle
    p^\beta,\xi_{\sigma(s_\beta+1)}-
\xi_{\sigma(s_\beta+2)}+\dots - \xi_{\sigma(s_\beta+N_\beta)}\rangle}\bigg)
\nonumber
\\
& \times \frac{1}{(8\pi)^{D/2} S_{(N_1,\dots,N_B)}}\;
\tilde{G}\Big(\underbrace{\tfrac{\|p^1\|^2}{2\mu^2},\dots,
\tfrac{\|p^1\|^2}{2\mu^2}}_{N_1}\big|\dots\big|
\underbrace{\tfrac{\|p^B\|^2}{2\mu^2},\dots,
\tfrac{\|p^B\|^2}{2\mu^2}}_{N_B}\Big).
\label{Schwinger-final}
\end{align}
This shows that out of the Osterwalder-Schrader axioms 
\cite{Osterwalder:1973dx, Osterwalder:1974tc}, 
\emph{Euclidean invariance} and
\emph{symmetry} are automatically fulfilled, whereas \emph{clustering} does not
hold. The remaining section addresses \emph{reflection
positivity} for the Schwinger 2-point function $S_c(\mu \xi_1,\mu \xi_2)$.

\subsection{Reflection positivity of Schwinger 2-point 
function for $D=6$}

It was proved in 
\cite{Grosse:2013iva} that the Schwinger 2-point function 
$S_c(\mu\xi_1,\mu\xi_2)$ given by (\ref{Schwinger-final}) is reflection
positive iff $x\mapsto 
\tilde{G}(x,x)
=G((2x+1)^2,(2x+1)^2)$ is a Stieltjes function, i.e.\ there exists a
positive measure $\varrho$ on $\mathbb{R}_+$ with 
$G((2x+1)^2,(2x+1)^2)=\int_0^\infty \frac{dt \varrho(t)}{t+2x}$. 
From (\ref{GXN}) and a combination of (\ref{MS-solution}) and 
(\ref{MS-solution-b}) we have in $D=6$ dimensions
\begin{align}
G(X,X)&=2W'(X) 
\label{GXX}
\\
&= \frac{\sqrt{1{+}c}}{\sqrt{X{+}c}}
+\frac{1}{2\sqrt{X{+}c}}
\int_1^\infty \!\!
\frac{dT\; \rho(T)}{\sqrt{T+c}}
\Big( \frac{1}{(\sqrt{1{+}c} + \sqrt{T{+}c})^2}
- \frac{1}{(\sqrt{X{+}c} + \sqrt{T{+}c})^2}\Big),
\nonumber
\end{align}
where $X=(2x+1)^2$ and $c \in {]{-}1,0]}$ for $\tilde{\lambda} \in
\mathbb{R}$. Already at this point we can state that reflection
positivity is impossible for $\tilde{\lambda} \in
\mathrm{i}\mathbb{R}\Leftrightarrow c>0$. Namely, $0=\sqrt{X{+}c}=
\sqrt{2x+1+\mathrm{i}\sqrt{c}}\sqrt{2x+1-\mathrm{i}\sqrt{c}}$ has
solutions -- hence (\ref{GXX}) a pole or end point of a branch cut -- 
\emph{outside} the real axis. This contradicts holomorphicity of Stieltjes
functions on $\mathbb{C}\setminus {]{-}\infty,0]}$.

\begin{figure}[h]
\parbox{12cm}{\includegraphics[width=12cm,bb=0 2 259 188,
viewport=0 2 259 188]{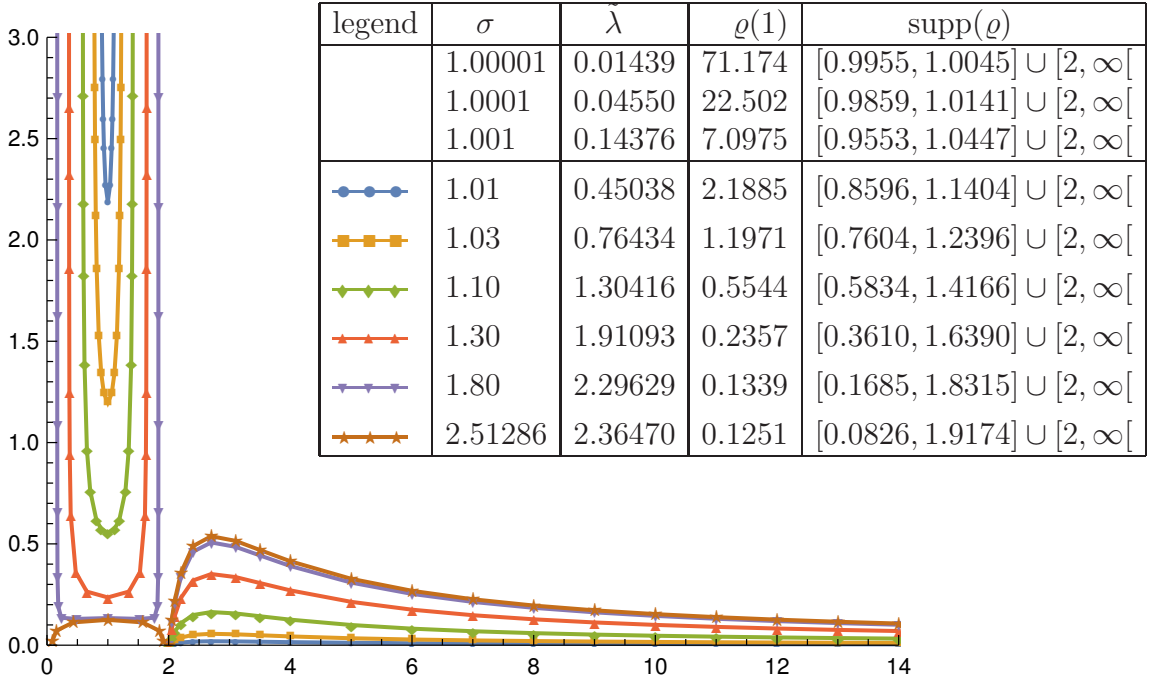}}
\hspace*{-8cm}\parbox[c]{5cm}{
$\begin{array}{|l|l|l|l|l|}
\hline
\text{legend} & \quad\sigma & \quad\tilde{\lambda} & \quad \varrho(1) & 
\qquad\quad \text{supp}(\varrho) 
\\ \hline
& 1.00001 & 0.01439 &71.174 &[0.9955, 1.0045]  \cup {[2,\infty[}
\\
& 1.0001 & 0.04550 & 22.502 &[0.9859, 1.0141]  \cup {[2,\infty[}
\\
& 1.001 & 0.14376 & 7.0975 &[0.9553, 1.0447] \cup {[2,\infty[}
\\ \hline
{\color{math1}\begin{picture}(10,5)
\thicklines\put(0,1){\line(1,0){10}}
\multiput(0.5,0)(3.5,0){3}{\mbox{\small$\bullet$}}
  \end{picture}}
& 1.01 & 0.45038 & 2.1885 & [0.8596, 1.1404] \cup {[2,\infty[}
\\
{\color{math2}\begin{picture}(10,5)
\thicklines\put(0,1){\line(1,0){10}}
\multiput(0.5,.2)(3.5,0){3}{\mbox{\tiny$\blacksquare$}}
  \end{picture}}
& 1.03 & 0.76434 & 1.1971 & [0.7604, 1.2396] \cup {[2,\infty[}
\\
{\color{math3}\begin{picture}(10,5)
\thicklines\put(0,1){\line(1,0){10}}
\multiput(0.5,.2)(3.5,0){3}{\mbox{\tiny$\blacklozenge$}} 
\end{picture}}
& 1.10 & 1.30416  & 0.5544 & [0.5834, 1.4166 ] \cup {[2,\infty[} 
\\
{\color{math4}\begin{picture}(10,5)
\thicklines\put(0,1){\line(1,0){10}}
\multiput(0.5,.4)(3.5,0){3}{\mbox{\tiny$\blacktriangle$}} 
\end{picture}}
& 1.30 & 1.91093 & 0.2357 & [0.3610, 1.6390 ] \cup {[2,\infty[} 
\\
{\color{math5}\begin{picture}(10,5)
\thicklines\put(0,1){\line(1,0){10}}
\multiput(0.5,.3)(3.5,0){3}{\mbox{\tiny$\blacktriangledown$}} 
\end{picture}}
& 1.80 & 2.29629 & 0.1339 & [0.1685, 1.8315 ] \cup {[2,\infty[} 
\\
{\color{math6}\begin{picture}(10,5)
\thicklines\put(0,1){\line(1,0){10}}
\multiput(0.5,.4)(3.5,0){3}{\mbox{\tiny$\bigstar$}} 
\end{picture}}
& 2.51286 & 2.36470 & 0.1251 & [0.0826, 1.9174]\cup {[2,\infty[} 
\\ \hline
\end{array}$
\vspace*{3cm}
}
\caption{Stieltjes measure $\varrho(t)$ of the diagonal 2-point 
function for $D=6$ and selected coupling constants. \label{fig1}
}\end{figure}

So let $\tilde{\lambda} \in \mathbb{R}$. For linearly spaced eigenvalues with 
$\rho(T)=\frac{\tilde{\lambda}^2(\sqrt{T}-1)^2}{4\sqrt{T}}$ we either 
evaluate (\ref{GXX}) or better differentiate (\ref{WX-D=6}) to 
\begin{align}
G(X,X) &= \frac{\sqrt{1+c}}{\sqrt{X+c}}
+  \frac{\tilde{\lambda}^2 (\sqrt{X}+1)(\sqrt{X}-1)}{
4 (\sqrt{X})^2 }\Big\{\frac{1}{\sqrt{X+c}}
\Big(\frac{1}{\sqrt{1+c}+\sqrt{X+c}}-1\Big)
\label{2W'X}
\\*
& +  \frac{1}{\sqrt{X}}
\big(\log (\sqrt{X}+\sqrt{X+c})-\log (\sqrt{X}\sqrt{1+c}+\sqrt{X+c})
+\log(1+\sqrt{X})\big)\Big\}.
\nonumber
\end{align}
The following is the deepest result of this paper:
\begin{thm}
\label{thm:positivity}
  The diagonal 2-point function of the renormalised 6-dimensional
  Kontsevich model $\Phi^3_6$ is, for linearly spaced eigenvalues of
  $E$, real $\tilde{\lambda}$ and in large-$(\mathcal{N},V)$ limit, a
  Stieltjes function, i.e.\ the Stieltjes transform 
$\tilde{G}(x,x)= \int_0^\infty \frac{\varrho(t) dt}{t+2x}$
of a positive measure $\varrho$. This Stieltjes measure $\varrho(t)$ 
has support 
$[1-\sqrt{-c},1+\sqrt{-c}] \cup [2,\infty[$ consisting of an 
isolated region near $t=1$ and the unbounded interval $t\geq 2$.
The precise relation is 
\begin{align}
\tilde{G}\Big(\frac{p^2}{2\mu^2},\frac{p^2}{2\mu^2}\Big)
&= \frac{\tilde{\lambda}^2}{4\pi(\sigma^2-1)}  \int_0^\pi\!\!\!
d\phi \;
\frac{\begin{array}{l}
\big\{2 \frac{\log (1+\sigma)}{\sigma}{-}1
+\sigma(\sigma{-}1)\tan^2 \phi \\
 -  \tan \phi \big(1{+}\sigma^2\tan^2 \phi\big) 
\big({\arctan\displaylimits_{[0,\pi]} (\sigma \tan \phi)}- \phi\big)\big\}
\end{array}
}{
1-\frac{\sqrt{\sigma^2-1}}{\sigma} \cos \phi +\frac{p^2}{\mu^2}}
\nonumber
\\*
& + 
\frac{\tilde{\lambda}^2}{4} \int_2^\infty\!\!\! dt \;
\frac{t(t-2)/(t-1)^3}{
t+\frac{p^2}{\mu^2}},
\end{align}
where $\sigma:=\frac{1}{\sqrt{1+c}}\in
[1,-2W_{-1}(-\frac{1}{2\sqrt{e}})-1]$ is the inverse solution of
$\tilde{\lambda}^2=\frac{4(\sigma^2-1)}{\sigma^2-2\sigma+2\log(1+\sigma)}
 \in [1,\frac{8W_{-1}(-\frac{1}{2\sqrt{e}})}{1+2W_{-1}(-\frac{1}{2\sqrt{e}})}] $.
Here, $W_{-1}(z)$ for $z\in [-\frac{1}{e},0]$ is the lower real 
branch of the Lambert-W function. 
\end{thm}
It should be possible to verify the claim directly.  The function
$\frac{\tan^n \phi}{a + b \cos\phi}$ has a known primitive so that
integration by parts does the job. 
Below we explain how we obtained the result. Figure~\ref{fig1} shows a plot of
the Stieltjes measure $\varrho(t)$ for various values of $\sigma$.

\subsection{Identification of the Stieltjes measure 
for $D=6$}

As long as $x>0$ the formula (\ref{2W'X}) can be taken literally.
But for discussing the Stieltjes property we have to extend it 
to complex $x$. Here we already made a choice for the logarithms:
When arranging them as in (\ref{2W'X}) we understand that $\log z$ has
a branch cut along the negative real axis and we choose the standard branch 
$\mathrm{Im}(\log z)\in {]0,\pi[}$ for $\mathrm{Im}(z)>0$ and 
$\mathrm{Im}(\log z)\in {]{-}\pi,0[}$ for $\mathrm{Im}(z)<0$. 
Next, the only reasonable interpretation that applies to 
$c \in {]{-}1,0]}$ is 
$\sqrt{X+c}:=\sqrt{(2x+1)+\sqrt{-c}}\sqrt{(2x+1)-\sqrt{-c}}$
and $\sqrt{X}:=2x+1$. This shows that 
$x\mapsto 2W'((2x+1))$ is holomorphic on $\mathbb{C}\setminus
{]{-}\infty,0[}$ with branch cut along parts of the negative real axis. 
Such holomorphicity is one of the characterising properties of 
Stieltjes functions. If we knew that $\tilde{G}(x,x)$ is Stieltjes, then 
the measure is recovered from the inversion formula
\begin{align}
\tilde{G}(x,x)= \int_0^\infty \frac{\varrho(t) dt}{t+2x}
\qquad\Rightarrow\qquad
\varrho(t)=\frac{1}{\pi} \mathrm{Im}
\big(\tilde{G}(-\tfrac{t}{2}-\mathrm{i}\epsilon,
-\tfrac{t}{2}-\mathrm{i}\epsilon)\big).
\label{Stieltjes-inversion}
\end{align}

To determine the nature of the branch cut we set 
$2x=-(t+\mathrm{i}\epsilon)$, $t>0$, and note
\begin{align}
\sqrt{X+c}\Big|_{\sqrt{X}=1-t-\mathrm{i}\epsilon}
=\left\{ \begin{array}{cl}
\sqrt{(1-t)^2+c} -\mathrm{i}\epsilon & \text{for } 
0\leq t < 1-\sqrt{-c} , 
\\
(-\mathrm{i}) \sqrt{-c-(1-t)^2}
& \text{for } 
1-\sqrt{-c} < t < 1+\sqrt{-c} , 
\\
-\sqrt{(t-1)^2+c}-\mathrm{i}\epsilon & \text{for } 
t \geq 1+\sqrt{-c}.
\end{array}\right.
\label{X+c}
\end{align}

We have to distinguish the three cases in (\ref{2W'X}):
\begin{enumerate}
\item{}[$0< t < 1-\sqrt{-c}$]:\qquad 
We have $\sqrt{X},\sqrt{X+c}>0$, hence
$\mathrm{Im}(2W'(X)) \Big|_{{\sqrt{X}=1-t-\mathrm{i}\epsilon
\atop 0\leq t \leq 1+\sqrt{-c}}} =0$.

\item {}[$1-\sqrt{-c} < t < 1+\sqrt{-c}$]:\qquad
Along the branch cut of 
$\sqrt{X+c}$, where this square root has strictly negative imaginary
part, the logarithms in (\ref{2W'X}) are well-defined, and we have
\begin{align}
&\mathrm{Im}(2W'(X)) \Big|_{{\sqrt{X}=1-t-\mathrm{i}\epsilon
\atop 1-\sqrt{-c} \leq t \leq 1+\sqrt{-c}}}
\label{2W'Xim}
\\
&= \frac{\sqrt{1+c}}{\sqrt{-c-(1-t)^2}}
-  \frac{\tilde{\lambda}^2 (\sqrt{1+c} - t(2-t))}{
4 (1-t)^2 \sqrt{-c-(1-t)^2}}
\nonumber
\\
& -  \frac{\tilde{\lambda}^2 t(2-t)}{4 (1-t)^3 }
\Big\{
\arctan\displaylimits_{[0,\pi]}\Big(\frac{\sqrt{-c-(1-t)^2}}{
(1-t)\sqrt{1+c}}\Big)
-\arctan\displaylimits_{[0,\pi]}\Big(\frac{\sqrt{-c-(1-t)^2}}{1-t}\Big)
\Big\}.
\nonumber
\end{align}
Positivity will be discussed below.

\item{}[$t > 1+\sqrt{-c}$]:\qquad 
The negative roots $\sqrt{X},\sqrt{X+c} \in
\mathbb{R}_--\mathrm{i}\epsilon$ 
are selected so that  
$\log (\sqrt{X}+\sqrt{X+c})-\log (\sqrt{X}\sqrt{1+c}+\sqrt{X+c})$ is
real. But $\log(1+\sqrt{X})=\log(2-t-\mathrm{i}{\epsilon})$ develops
an imaginary part for $t > 2$ (recall that $\log$ is the standard branch):
\begin{align}
&\mathrm{Im}(2W'(X)) \Big|_{{\sqrt{X}=1-t-\mathrm{i}\epsilon
\atop t \geq 1+\sqrt{-c}}}
= \chi_{[2,\infty[} 
 \frac{\tilde{\lambda}^2 \pi t (t-2)}{4(t-1)^3 } ,
\label{2W'Xim-2}
\end{align}
where $\chi_{[2,\infty[} $ is the characteristic function of
${[2,\infty[}$. The function (\ref{2W'Xim-2}) is manifestly non-negative and, 
remarkably, depends on 
$\tilde{\lambda}$ only via the global prefactor $\tilde{\lambda}^2$,
but not on $c(\tilde{\lambda})$.

\end{enumerate}

Case 2 needs careful discussion. For the recognition of $2W'(X)$ 
as a Stieltjes function we need $\mathrm{Im}(2W'(X))\geq 0$ for 
$\sqrt{X}=1-t-\mathrm{i}\epsilon$. 
We find it convenient to introduce in (\ref{2W'Xim}) 
and (\ref{W1=1}) the substitution
\begin{align}
\sqrt{-c}=\cos \psi,\quad
\sqrt{1+c}=\sin\psi,\quad
1-t=\cos \psi\cos\phi.
\end{align}
In these variables we have, after extracting 
a common prefactor $\tilde{\lambda}^2$, 
\begin{align}
\underbrace{\mathrm{Im}(2W'(X))}_{\sqrt{X}=\cos \psi\cos\phi-\mathrm{i}\epsilon}
\label{2W'Xim-psi}
&=
\frac{\tilde{\lambda}^2 }{4 \sin \phi \cos^3 \psi }\Big\{
\sin^2 \psi \Big(2 \sin \psi \log (1+\frac{1}{\sin \psi})-1\Big)
\\[-2ex]
& +(1-\sin \psi) \tan^2 \phi  -  \tan \phi (\sin^2 \psi +\tan^2 \phi) 
\Big(\arctan\displaylimits_{[0,\pi]} 
\Big(\frac{\tan \phi}{\sin \psi}\Big)- \phi\Big)
\Big\}.
\nonumber
\end{align}
For $\phi\to 0$ we see that positivity requires 
$\log (1+\frac{1}{\sin \psi})\geq \frac{1}{2\sin \psi}$
or 
\begin{align}
\psi &\geq \arcsin\Big(\frac{1}{-2W_{-1}(-\frac{1}{2\sqrt{e}})-1}\Big) =
0.409284\dots, \qquad
\end{align}
where $W_{-1}(z)$ is the lower branch of the Lambert-W function.
This gives a bound on the coupling constant
\begin{align}
\sqrt{1+c}&=\sin \psi \geq  \frac{1}{-2W_{-1}(-\frac{1}{2\sqrt{e}})-1}
=0.397953\dots.,\qquad |\lambda| \leq  2.3647\dots.
\end{align}
Coincidently, this critical value agrees with the critical
value where $\frac{d\tilde{\lambda}^2(c)}{dc}=0$. In other words, we
have positivity precisely on the interval where 
$c\mapsto \tilde{\lambda}^2(c)$ is bijective.

The other critical value to discuss is $\phi \to \frac{\pi}{2}$, or $t\to 1$, 
where $\tan\phi$ becomes singular. The series expansion
yields
\begin{align}
&\mathrm{Im}(2W'(X)) \Big|_{{\sqrt{X}=1-t-\mathrm{i}\epsilon \atop
t\to 1}}
\label{2W'Xim-ser}
\\*
&=
\frac{\tilde{\lambda}^2 }{12 \cos^3 \psi}\Big(
 (1 - 6 \sin^2\psi + 2 \sin^3 \psi + 
    6 \sin^3 \psi \log\Big(1+\frac{1}{\sin \psi}\Big)\Big) 
+\mathcal{O}((1-t)^2),
\nonumber
\end{align}
which is manifestly positive. Inserting (\ref{2W'Xim-psi}) and
(\ref{2W'Xim-2}) into 
the invesion formula (\ref{Stieltjes-inversion}) and taking the
Jacobian of $dt=\cos\psi \sin \phi \,d\phi$ into account 
we arrive at Theorem~\ref{thm:positivity}.

\subsection{Reflection positivity of 2-point 
function for $D=4$}

\begin{figure}[h]
\parbox{11cm}{\includegraphics[width=11.5cm,bb=0 2 259 188,
viewport=0 2 259 190]{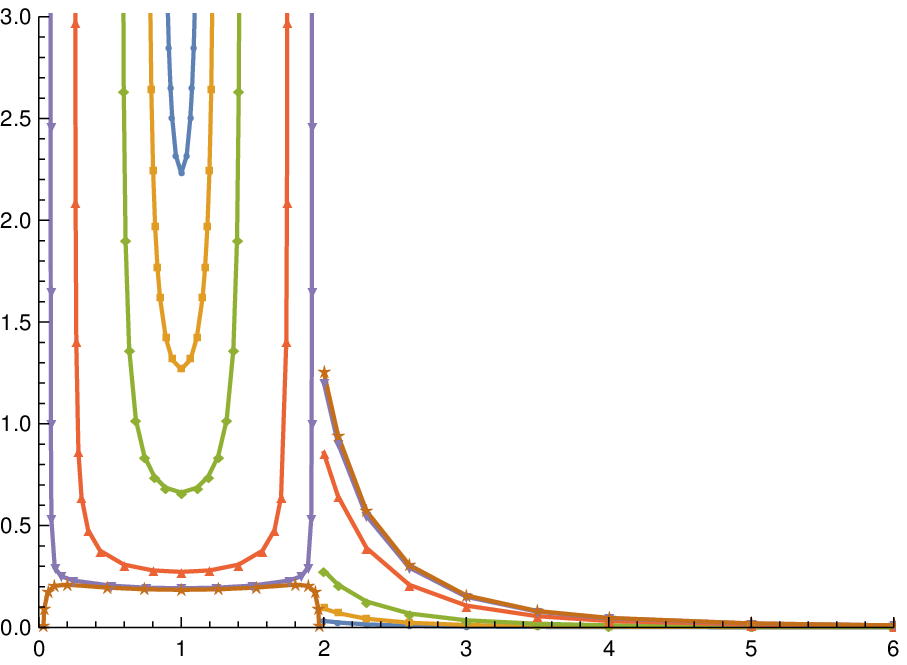}}
\hspace*{-6.5cm}\parbox[c]{5cm}{
$\begin{array}{|l|l|l|l|l|}
\hline
\text{legend} & \quad\sigma & \quad\tilde{\lambda} & \quad \varrho(1) & 
\qquad\quad \text{supp}(\varrho) 
\\ \hline
& 1.00001 & 0.00571 &71.176 &[0.9955, 1.0045]  \cup {[2,\infty[}
\\
& 1.0001 & 0.01805 & 22.506 &[0.9859, 1.0141]  \cup {[2,\infty[}
\\
& 1.001 & 0.05704 & 7.1114 &[0.9553, 1.0447] \cup {[2,\infty[}
\\ \hline
{\color{math1}\begin{picture}(10,5)
\thicklines\put(0,1){\line(1,0){10}}
\multiput(0.5,0)(3.5,0){3}{\mbox{\small$\bullet$}}
  \end{picture}}
& 1.01 & 0.17907 & 2.2315 & [0.8596, 1.1404] \cup {[2,\infty[}
\\
{\color{math2}\begin{picture}(10,5)
\thicklines\put(0,1){\line(1,0){10}}
\multiput(0.5,.2)(3.5,0){3}{\mbox{\tiny$\blacksquare$}}
  \end{picture}}
& 1.03 & 0.30525 & 1.2676 & [0.7604, 1.2396] \cup {[2,\infty[}
\\
{\color{math3}\begin{picture}(10,5)
\thicklines\put(0,1){\line(1,0){10}}
\multiput(0.5,.2)(3.5,0){3}{\mbox{\tiny$\blacklozenge$}} 
\end{picture}}
& 1.10 & 0.52847  & 0.6621 & [0.5834, 1.4166 ] \cup {[2,\infty[} 
\\
{\color{math4}\begin{picture}(10,5)
\thicklines\put(0,1){\line(1,0){10}}
\multiput(0.5,.4)(3.5,0){3}{\mbox{\tiny$\blacktriangle$}} 
\end{picture}}
& 1.50 & 0.92552 & 0.2726 & [0.2546, 1.7454 ] \cup {[2,\infty[} 
\\
{\color{math5}\begin{picture}(10,5)
\thicklines\put(0,1){\line(1,0){10}}
\multiput(0.5,.3)(3.5,0){3}{\mbox{\tiny$\blacktriangledown$}} 
\end{picture}}
& 2.50 & 1.09666 & 0.1922 & [0.0835, 1.9165 ] \cup {[2,\infty[} 
\\
{\color{math6}\begin{picture}(10,5)
\thicklines\put(0,1){\line(1,0){10}}
\multiput(0.5,.4)(3.5,0){3}{\mbox{\tiny$\bigstar$}} 
\end{picture}}
& 3.92155 & 1.12027 & 0.1843 & [0.0331, 1.9669]\cup {[2,\infty[} 
\\ \hline
\end{array}$
\vspace*{3cm}
}
\caption{Stieltjes measure of the diagonal 2-point function for $D=4$ and 
selected coupling constants. \label{fig2}
}\end{figure}

For linearly spaced eigenvalues in $D=4$ dimensions, hence summation measure 
$\rho(T)=\tilde{\lambda}^2 \frac{\sqrt{T-1}}{\sqrt{T}}$, we evaluate 
(\ref{MS-solution-b4}) to
\begin{align}
\tilde{\lambda}^2 = \frac{1-\sqrt{1+c}}{1-\sqrt{1+c}
  \log(1+\frac{1}{\sqrt{1+c}})} .
\end{align}
Next, (\ref{MS-solution}) with $Z=1$ is evaluated to
\begin{align}
\label{2W'-D=4}
2W'(X)&= \frac{1}{\sqrt{X+c}}\Big(1- \frac{\tilde{\lambda}^2}{2}
\int_1^\infty
\frac{dT\;(\sqrt{T}-1)}{\sqrt{T}\sqrt{T+c}(\sqrt{X+c}+\sqrt{T+c})^2}\Big)
\\
\nonumber
&= \frac{1}{\sqrt{X+c}} 
- \frac{\tilde{\lambda}^2}{\sqrt{X}^2}
+ \frac{\tilde{\lambda}^2(\sqrt{1+c}-1)}{\sqrt{X}^2\sqrt{X+c}}
\nonumber
\\
& + \frac{\tilde{\lambda}^2}{\sqrt{X}^3} 
\Big(\log(\sqrt{X}+1)+\log(\sqrt{X}+\sqrt{X+c})-
\log(\sqrt{X}\sqrt{1+c}+\sqrt{X+c})\Big).
\nonumber
\end{align}
The same discussion as for $D=6$ gives:
\begin{thm}
  The diagonal 2-point function of the renormalised 4-dimensional
  Kontsevich model $\Phi^3_4$ is, for linearly spaced eigenvalues of
  $E$, real $\tilde{\lambda}$ and in large-$(\mathcal{N},V)$ limit, a
  Stieltjes function. The Stieltjes measure $\varrho(t)$ 
has support 
$[1-\sqrt{-c},1+\sqrt{-c}] \cup [2,\infty[$ consisting of an 
isolated region near $t=1$ and the unbounded interval $t\geq 2$.
The precise relation is 
\begin{align}
\tilde{G}\Big(\frac{p^2}{2\mu^2},\frac{p^2}{2\mu^2}\Big)
&= \frac{\tilde{\lambda}^2\sigma^2 }{\pi(\sigma^2-1)}  \int_0^\pi\!\!\!
d\phi \;
\frac{\begin{array}{l}
\big\{\frac{(1+\sigma)}{\sigma} (1-\frac{\log (1+\sigma)}{\sigma})
\\
-(1 {+} \tan^2\phi)\big(1-\frac{1}{\sigma} -  \tan \phi 
({\arctan\displaylimits_{[0,\pi]} (\sigma \tan \phi)}- \phi)\big)\big\}
\end{array}
}{
1-\frac{\sqrt{\sigma^2-1}}{\sigma} \cos \phi +\frac{p^2}{\mu^2}}
\nonumber
\\*
& +
\tilde{\lambda}^2 \int_2^\infty\!\!\! dt \;
\frac{1/(t-1)^3}{
t+\frac{p^2}{\mu^2}},
\end{align}
where $\sigma:=\frac{1}{\sqrt{1+c}}\in
[1,-\frac{2}{W_0(-\frac{2}{e^2})}-1]$ is the inverse solution of
$\tilde{\lambda}^2=
\frac{\sigma -1}{\sigma - \log(1+\sigma)}
\in [0, \frac{2}{2+W_0(-\frac{2}{e^2})}]$.
Here, $W_0(z)$ for $z \geq -\frac{1}{e}$ is the upper real 
branch of the Lambert-W function. 
\end{thm}
\noindent
Positivity of the measure at $\phi\in \{0,\pi\}$ leads to 
the same condition 
$2 (1+\sigma) -(1+\sigma)\log (1+\sigma)\geq 2$ (solved in terms of
Lambert-W) that restricts the bijectivity region of $c\mapsto
\tilde{\lambda}^2(c)$. Figure~\ref{fig2} shows a plot of the
Stieltjes measure for various values of $\sigma$.

\subsection{Two-point function for $D=2$}

We had already pointed out in \cite{Grosse:2016pob} that the
$\Phi^3_2$-model is \emph{not} reflection positive. In this subsection
we show what goes wrong compared with $D=4$ and $D=6$. Starting point
is (\ref{MS-solution}) which evaluates for the measure 
$\rho(T)=\frac{2\tilde{\lambda}^2}{\sqrt{T}}$ to 

\begin{align}
2W'(X)&= \frac{1}{\sqrt{X+c}}\Big(1- \tilde{\lambda}^2 
\int_1^\infty
\frac{dT}{\sqrt{T}\sqrt{T+c}(\sqrt{X+c}+\sqrt{T+c})^2}\Big)
\label{2W'X:D=2}
\\*
&= \frac{1}{\sqrt{X+c}} + \frac{2\tilde{\lambda}^2}{\sqrt{X}^2
(\sqrt{X+c}+\sqrt{1+c})\sqrt{1+c}} + \frac{2\tilde{\lambda}^2}{\sqrt{X}^2
\sqrt{X+c}} \Big(1-\frac{1}{\sqrt{1+c}}\Big)
\nonumber
\\*
& - \frac{2\tilde{\lambda}^2}{\sqrt{X}^3} 
\Big(\log(\sqrt{X}+1)+\log(\sqrt{X}+\sqrt{X+c})-
\log(\sqrt{X}\sqrt{1+c}+\sqrt{X+c})\Big).
\nonumber
\end{align}
\emph{It is the opposite sign of the last 
line in (\ref{2W'X:D=2}) compared with 
(\ref{2W'-D=4}) which lets the scattering measure supported at $t>2$
arise with the wrong sign!} 
In addition there is an atomic measure
from the second term on the rhs of (\ref{2W'X:D=2}):
Near $t=2$ we have 
\begin{align*}
&\mathrm{Im}\Big( \frac{2\tilde{\lambda}^2/\sqrt{1+c}}{\sqrt{X}^2
(\sqrt{X+c}+\sqrt{1+c})}\Big)
\stackrel{\sqrt{X}=-\frac{t}{2}-\mathrm{i}\epsilon}{\longrightarrow}
\mathrm{Im}\Big(
\frac{2\tilde{\lambda}^2/\sqrt{1+c}}{(t-1)^2
(-\sqrt{(t-1)^2+c}+\sqrt{1+c} - \mathrm{i}\epsilon)}
\Big)
\\
&=
\frac{2\pi \tilde{\lambda}^2}{(t-1)^2\sqrt{1+c}}
\delta\Big(\sqrt{(t-1)^2+c}-\sqrt{1+c}\Big)
= 2\pi \tilde{\lambda}^2\delta(t-2).
\end{align*}
Adding also the measure on $[1-\sqrt{-c},1+\sqrt{-c}]$
we obtain the representation:
\begin{prop}
  The diagonal 2-point function of the renormalised 2-dimensional
  Kontsevich model $\Phi^3_2$ has, for linearly spaced eigenvalues of
  $E$, real $\tilde{\lambda}$ and in large-$(\mathcal{N},V)$ limit, 
 an integral representation
\begin{align}
\tilde{G}\Big(\frac{p^2}{2\mu^2},\frac{p^2}{2\mu^2}\Big)
&= \frac{2\tilde{\lambda}^2\sigma^2 }{\pi(\sigma^2-1)}  \int_0^\pi\!\!\!
d\phi \;
\frac{\begin{array}{l}
\big\{(1+\sigma) \frac{\log (1+\sigma)}{\sigma}{-}
\frac{\sigma (1 {+} \tan^2\phi)^2}{1{+} \sigma^2 \tan^2\phi}
\\
+(1 {+} \tan^2\phi)\big(1-  \tan \phi 
({\arctan\displaylimits_{[0,\pi]} (\sigma \tan \phi)}- \phi)\big)\big\}
\end{array}
}{
1-\frac{\sqrt{\sigma^2-1}}{\sigma} \cos \phi +\frac{p^2}{\mu^2}}
\nonumber
\\*
& -
2\tilde{\lambda}^2 \int_2^\infty\!\!\! dt \;
\frac{1/(t-1)^3}{
t+\frac{p^2}{\mu^2}}+\frac{\tilde{\lambda}^2}{\frac{p^2}{\mu^2}+1}.
\end{align}
Therefore, $p^2 \mapsto 
\tilde{G}(\frac{p^2}{2\mu^2},\frac{p^2}{2\mu^2}) $ 
is not a Stieltjes function!
\end{prop}

\section{Summary and outlook}

We extended our previous work \cite{Grosse:2016pob} (on $D = 2$) to
give an exact solution of the $\Phi^3_D$ large-$\mathcal{N}$ matrix
model in $D = 4$ and $D = 6$ dimensions. Induction proofs and the
difficult combinatorics were unchanged compared with $D = 2$, but the
renormalisation, performed in the manner of Wolfhart Zimmermann, was
much more involved. The main lesson is that our method is powerful
enough to handle just renormalisable models with running coupling
constant where a perturbative approach is plagued by overlapping
divergences and renormalon problem. None of these perturbative
artefacts arises: the exact renormalised correlation functions are
analytic in the renormalised $\Phi^3$-coupling constant. Although the
bare (and real) $\Phi^3_6$-coupling constant diverges (positive
$\beta$-function), the exact solution does not develop a Landau pole.

The deepest result established in this paper is the completely
unexpected proof that the Schwinger 2-point function arising from the
large-deformation limit of the $\Phi^3_D$-model on noncommutative
Moyal space is reflection positive in $D=4$ and $D=6$ dimensions, but
not in $D=2$. This result relied heavily on the explicit knowledge of
the $\Phi^3_D$-matrix correlation functions which allowed us to
perform the analytic continuation to the complex plane. Consequently,
this sector of the theory defines unambiguously a Wightman 2-point
function of a true relativistic quantum field theory
\cite{Streater:1964??} in $D\in \{4,6\}$ dimensions. 

We explicitly computed the Stieltjes measure of the
Euclidean quantum field theory, which is the same as the
K\"all\'en-Lehmann mass spectrum \cite{Kallen:1952zz,  Lehmann:1954xi}
of the Wightman theory. The mass
shell around $|p|^2=\mu^2$ is not sharp but fuzzy with a non-zero
width which depends on the coupling constant. In addition there is a
scattering spectrum starting at $|p|^2=2\mu^2$ and ranging up to
$\infty$. The beginning at $|p|^2=2\mu^2$ and not at $|p|^2=4\mu^2$ is
strange. It essentially means that the theory, although in
dimension $D\in \{4,6\}$, behaves like a one-dimensional theory where
only the energy, and no momentum, is additive. This sounds somewhat
disappointing, but mathematical physics in one dimension
\cite{Lieb:2013??} is a rich subject! Of course it remains to be 
seen whether the interpretation as a scattering spectrum is correct. 
According to Aks \cite{Aks:1965??}, scattering in dimension $D\in \{4,6\}$
must be accompanied by particle production, which however is 
absent in integrable models. The clarification of this question, and of 
reflection positivity of $(N>2)$-point functions, are left 
for future investigation.

\section*{Acknowledgements}

\noindent 
We thank Prof.\ Klaus Sibold for inviting us to contribute a paper to
the memory of Prof.\ Wolfhart Zimmermann. R.W.\ would like to thank
the Faculty of Physics of the University of Vienna for support of and
hospitality during a visit where the initial steps of this work were
done. A.S.\ was supported by JSPS KAKENHI Grant Number 16K05138, and
R.W.\ by SFB 878.


\end{document}